\documentclass[11pt]{article}
\usepackage{amsmath,amssymb,exscale}
\usepackage{color,graphicx,epsfig}
\usepackage[utf8]{inputenc}
\usepackage{hyperref}
\usepackage{multirow}
\usepackage{subfigure}
\usepackage{cite}
\usepackage{epsfig}
\input epsf
\newcommand{\m}[1]{\marginpar{{\tiny *}} }

\begin{document}
\topmargin -1.0cm
\oddsidemargin 0cm
\evensidemargin 0cm

\thispagestyle{empty}

\vspace{40pt}

\begin{center}
\vspace{40pt}

\Large \textbf{Constraints on scalar and vector leptoquarks \\ from the LHC Higgs data}

\end{center}

\vspace{15pt}
\begin{center}
	{\bf Jian Zhang $^{a}$}\footnote{E-mail: zhangjianphy@aliyun.com}
	\ {\bf Chong-Xing Yue $^{a}$}\footnote{E-mail: cxyue@lnnu.edu.cn} \ {\bf Chun-Hua Li $^{a}$} and {\bf Shuo Yang $^{b}$}
	\\
	{$^a$Department of Physics, Liaoning Normal University, Dalian 116029, China \\
		$^b$Department of Physics, Dalian University, Dalian 116622, China }
\end{center}

\vspace{20pt}
\begin{center}
\textbf{Abstract}
\end{center}
\vspace{5pt} {\small \noindent
We study contributions of single scalar or vector leptoquark (LQ) to loop-induced Higgs processes, gluon fusion production ($g g \to h$) and $h \to \gamma \gamma$ decay, by analyzing the current Higgs data from the LHC Run I and II. Scalar LQ is studied in a model independent way, while the vector LQ $U_1(\mathbf{3},\mathbf{1},2/3)$ is discussed in the '4321' model. Constraints on the interactions of LQ and  Higgs boson are obtained. We provide a method to determine vacuum expectation values $\upsilon_3$ and $\upsilon_1$ of the new scalar fields $\Omega_3$ and $\Omega_1$ in the '4321' model via the combination of Higgs data and measurements of $R_{D^{(*)}}$ and $R_{K^{(*)}}$.
}

\noindent
\eject

%\tableofcontents

\newpage

\section{Introduction}
\label{intro}

In the last few years accumulated experimental results of semileptonic $B$-meson decays point to lepton flavour universality
violation (LFUV). In the case of flavor changing neutral current (FCNC) transition $b \to s \mu^+ \mu^-$, ratios $R_{K^{(*)}} = \frac{\mathcal{B}(\bar{B} \to K^{(*)} \mu^+ \mu^-)}{\mathcal{B}(\bar{B} \to K^{(*)} e^+ e^-)}$ measured by the LHCb collaboration are lower than the SM expectations by $\sim 2.6 \sigma$~\cite{Aaij:2014ora,Aaij:2017vbb,Bordone:2016gaq,Capdevila:2017ert}. For $b \to c \ell \nu$ ($\ell = e, \mu$) charged current case, measurements of $R_{D} = \frac{\mathcal{B}(\bar{B} \to D \tau^- \bar{\nu})}{\mathcal{B}(\bar{B} \to D \ell^- \bar{\nu})}$ and $R_{D^*} = \frac{\mathcal{B}(\bar{B} \to D^* \tau^- \bar{\nu})}{\mathcal{B}(\bar{B} \to D^* \ell^- \bar{\nu})}$ from experiments are higher than the SM expectations by $\sim 2.3 \sigma$ and $\sim 3.4 \sigma$, respectively~\cite{Aaij:2015yra,Huschle:2015rga,Sato:2016svk,Hirose:2016wfn,Lees:2012xj,Lees:2013uzd,HFAG2017,Fajfer:2012vx,Aoki:2016frl}. As popular candidates for explaining $B$-anomalies, leptoquarks (LQs) are extensively discussed in specific ultraviolet (UV) theories or model-independently (see, e.g,~\cite{Gripaios:2014tna,Georgi:2016xhm,Becirevic:2016yqi,Becirevic:2017jtw,Diaz:2017lit,Buttazzo:2017ixm,Guo:2017gxp,DiLuzio:2017vat,Calibbi:2017qbu,Blanke:2018sro,Fajfer:2018bfj,Matsuzaki:2018jui,Hati:2018fzc,Becirevic:2018afm,Crivellin:2018yvo,deMedeirosVarzielas:2018bcy,Azatov:2018kzb,DiLuzio:2018zxy,Faber:2018qon,Heeck:2018ntp,Angelescu:2018tyl,Balaji:2018zna,Watanabe:2018jhh,Schmaltz:2018nls,Bansal:2018nwp,Iguro:2018vqb,Fajfer:2018hbq,Fornal:2018dqn,DaRold:2018moy,deMedeirosVarzielas:2019lgb,Zhang:2019hth,Aydemir:2019ynb,Cata:2019wbu,Bhattacharya:2019olg,Adam:2019oes,Aebischer:2019mlg,Cornella:2019hct,Barbieri:2019zdz}).

LQs are hypothetical color-triplet bosons that carry both baryon and lepton numbers~\cite{Buchmuller:1986zs,Dorsner:2016wpm,Tanabashi:2018oca}. They naturally appear in many extensions of the Standard Model (SM) such as Pati-Salam model~\cite{Pati-Salam}, grand unification theories based on $SU(5)$~\cite{Georgi:1974sy} and $SO(10)$~\cite{Georgi:1974my}, extended technicolor~\cite{technicolor}, and compositeness~\cite{composite}. According to their properties under the Lorentz transformations, LQs can be either scalar (spin 0) or vector (spin 1). Several models suggest LQs mass of TeV-scale.

LQs can also couple to Higgs boson and considerably modify loop-induced Higgs processes, gluon fusion production ($ggF$) and $h \to \gamma \gamma$ decay, without appreciably changing kinematics of theses process. Scalar LQs interact with the Higgs boson at tree level via Higgs portal interactions. Their contributions to loop-induced Higgs processes can be studied model-independently~\cite{Dorsner:2016wpm,Dorsner:2015mja,Chang:2012ta}. Vector LQs, as gauge fields in full fledged models, make contributions to the loop processes that are sensitive to the gauge sector of the ultraviolet (UV) theories which they belong to. $ggF$ predominates the Higgs production processes at the LHC. And the LHC is sensitive to $h \to \gamma \gamma$ decay process. After the discovery of the 125 GeV Higgs boson by the ATLAS \cite{atlas-higgs} and CMS \cite{cms-higgs} experiments in 2012, precisely measuring properties of the Higss boson are then performed by the ATLAS and CMS experiments with LHC Run I and II data sets~\cite{Khachatryan:2016vau,ATLAS:2019slw,Sirunyan:2018koj}. Globally analyzing these measurements, in some sense, can guide us for LQs study.

Constraints on scalar LQs are obtained by Ref.~\cite{Dorsner:2016wpm} via analyzing Higgs data from the LHC Run I reported by the ATLAS and CMS collaborations~\cite{TheATLASandCMSCollaborations:2015bln,CMS:2015kwa}. We update these results via comprehensively analyzing Higgs data from the LHC Run I and II~\cite{Khachatryan:2016vau,ATLAS:2019slw,Sirunyan:2018koj}.

Since interactions between vector LQ and the Higgs boson as well as other gauge fields are sensitive to the UV theories which the LQ belongs to, the contributions of vector LQ  to the loop-induced Higgs processes should be studed in a specific model. Of particular note is that $U_1(\mathbf{3},\mathbf{1},2/3)$ with mass of several TeV performs quite well in explaining both anomalies of $R_{D^{(*)}}$ and $R_{K^{(*)}}$~\cite{Buttazzo:2017ixm}. In this article, we study $U_1(\mathbf{3},\mathbf{1},2/3)$ originating from a particular theory, namely the '4321' model. One of purposes of the model is to explain $B$-anomalies~\cite{DiLuzio:2017vat,DiLuzio:2018zxy}. Besides obtaining the constraints on the size of vector LQ interactions to the Higgs boson from current LHC Higgs data, we also provide a method to determine vacuum expectation values (VEVs) $\upsilon_3$ and $\upsilon_1$ of the new scalar fields $\Omega_3$ and $\Omega_1$ in the '4321' model via the combination of Higgs data and measurements of $R_{D^{(*)}}$ and $R_{K^{(*)}}$.

The article is organized as follows: we first review current Higgs data from the LHC Run I and II in Section~\ref{sec:2}. In Section~\ref{sec:3}, we model-independently study the contributions of single scalar LQ to loop-induced Higgs processes, $ggF$ production and $h \to \gamma \gamma$ decay. Contributions of the vector LQ $U_1$ to these loop processes are discussed in framework of the '4321' model in Section~\ref{sec:4}. In the same section, we also discuss the determination of VEVs $\upsilon_3$ and $\upsilon_1$ of this model. Finally, conclusions for this work are given in Section~\ref{sec:5}.

\section{The LHC Higgs data}
\label{sec:2}

The discovery of the 125 GeV Higgs boson by the ATLAS~\cite{atlas-higgs} and CMS~\cite{cms-higgs} experiments in 2012 is one of the greatest achievements in the history of particle physics. Precise measurements of the Higss boson properties are then performed by these experiments. At the LHC, only products of cross sections and branching fractions are measured.
In the narrow-width approximation, the signal cross section of an individual channel, e.g. $\sigma (gg \to H \to \gamma\gamma)$, can be factorized as \cite{Heinemeyer:2013tqa}
\begin{eqnarray}
\sigma (gg \to H \to \gamma\gamma) &=& \frac{\sigma_{ggF} \cdot \Gamma^{\gamma\gamma}}{\Gamma_h} \nonumber \\
&=& (\sigma_{ggF} \cdot B^{\gamma\gamma})_{SM}\dfrac{\kappa_g^2 \cdot \kappa_{\gamma}^2}{\kappa_h^2},
\end{eqnarray}
where $\sigma_i$ and $\Gamma^j$ represent measured values of $i \to h$ production and $h \to j$ decay, respectively, and $\sigma^{SM}_i$ and $\Gamma^j_{SM}$ are their SM expectations, $\kappa_i$ are the so called 'coupling modifiers' defined as $\kappa^2_i = \sigma_i/\sigma^{SM}_{i}$ or $\kappa^2_i = \Gamma^i/\Gamma^i_{SM}$ ( all $\kappa_i$ values equal unity in the SM ), and $\Gamma_h$ denotes the total width of the Higgs boson.

In 2016, the ATLAS and CMS collaborations reported measurements of the Higgs boson production and decay rates as well as constraints on its couplings to vector bosons and fermions by using the LHC Run I data recorded in 2011 and 2012~\cite{Khachatryan:2016vau}. The integrated luminosities in each experiment are about 5 fb$^{-1}$ at $\sqrt{s} = 7$ TeV and 20 fb$^{-1}$ at $\sqrt{s} = 8$ TeV. The measurements are based on five main Higgs boson production processes (gluon fusion, vector boson fusion, and associated production with a $W$ or a $Z$ boson or pair of top quarks) and six decay modes ( $h \to ZZ, WW, \gamma \gamma, \tau \tau, bb \ {\rm and} \ \mu \mu$ ).

In 2019, the similar measurements are reported by the ATLAS and CMS collaborations via using the Run II data set recorded by the ATLAS detector during 2015, 2016 and 2017 with the integrated luminosity of 79.8 fb$^{-1}$ at $\sqrt{s} = 13$ TeV~\cite{ATLAS:2019slw} and the CMS detector in 2016 at $\sqrt{s} = 13$ TeV with the integrated luminosity of 35.9 fb$^{-1}$~\cite{Sirunyan:2018koj}, respectively.

The Higgs boson with mass of $m_h = 125.09$ GeV is assumed in all the above experimental analyses. These measurements normalized to the SM predictions are listed in Table~\ref{tab:measurements}. From Table~\ref{tab:measurements} we can see that measurements obtained by each experiment from the LHC Run I or Run II are precisely consistent within error with their SM predictions. This implies that NP properly lies in a scale much higher than the mass of Higgs boson, and new heavy particles carrying electric and colour charge may still be present in the loop-induced Higgs processes, $ggF$ production and $h \to \gamma \gamma$ decay, without appreciably changing kinematics of theses process~\cite{Chang:2012ta,Enkhbat:2013oba,Djouadi:2005gj,Carena:2012xa,Dorsner:2012pp,Agrawal:1999bk,Gori:2013mia}.

\begin{table}
	\caption{Best fit values of $\sigma(gg \to h \to ZZ)$, $\sigma_i/\sigma_{ggF}$ and $\mathcal{B}^f/\mathcal{B}^{ZZ}$ obtained from different experiments. The measurements are normalized to the SM predictions.}
	\label{tab:measurements}
	\centering
	\begin{tabular}{lr@{\hskip 0.4ex}lr@{\hskip 0.4ex}lr@{\hskip 0.4ex}l}\\
		\hline
		\hline\noalign{\smallskip}
		\multicolumn{1}{c}{\multirow{4}{*}{Measurements}} & \multicolumn{6}{c}{\multirow{2}{*}{Values}}\\
		\multicolumn{1}{c}{\multirow{6}{*}}& \\
		\cline{2-7}
		&\multicolumn{2}{c}{ATLAS \& CMS \cite{Khachatryan:2016vau}} & \multicolumn{2}{c}{ATLAS \cite{ATLAS:2019slw}} & \multicolumn{2}{c}{CMS \cite{Sirunyan:2018koj}} \\
		& \multicolumn{2}{c}{$\sqrt{s}=8$ TeV} & \multicolumn{2}{c}{$\sqrt{s}=13$ TeV} & \multicolumn{2}{c}{$\sqrt{s}=13$ TeV}\\
		\hline
		$\sigma_{ggF}\cdot B_{ZZ} $ & $1.16$&$ {}_{- \ 0.24}^{+ \
			0.26}$ & $1.13$&${}_{- \ 0.13}^{+ \ 0.13}$ & $1.07$&${}_{- \ 0.18}^{+ \ 0.20}$  \\
		$\sigma_{VBF}/\sigma_{ggF}$ & $1.33$&${}_{-\ 0.36}^{+\ 0.44}$  & $1.23$&${}_{-\ 0.27}^{+\ 0.32}$ & $0.6$&${}_{-\ 0.24}^{+\ 0.30}$  \\
		$\sigma_{WH}/\sigma_{ggF}$ & $0.84$&${}_{-\ 0.71}^{+\ 0.76}$ & $1.26$&${}_{-\ 0.45}^{+\ 0.59}$ & $2.19$&${}_{-\ 0.69}^{+\ 0.86}$  \\
		$\sigma_{ZH}/\sigma_{ggF}$ & $3.06$&${}_{-\ 1.48}^{+\ 1.84}$ & $1.01$&${}_{-\ 0.35}^{+\ 0.47}$ & $0.88$&${}_{-\ 0.27}^{+\ 0.34}$  \\
		$\sigma_{ttH+tH}/\sigma_{ggF}$ & $3.28$&${}_{-\ 1.02}^{+\ 1.15}$ & $1.20$&${}_{-\ 0.27}^{+\ 0.31}$ & $1.06$&${}_{-\ 0.27}^{+\ 0.34}$ \\
		$\mathcal{B}_{\gamma\gamma}/\mathcal{B}_{ZZ}$ & $0.81$&${}_{-\ 0.16}^{+\ 0.21}$ & $0.87$&${}_{-\ 0.12}^{+\ 0.14}$ & $1.14$&${}_{-\ 0.20}^{+\ 0.28}$  \\
		$\mathcal{B}_{WW}/\mathcal{B}_{ZZ}$ & $0.83$&${}_{-\ 0.16}^{+\ 0.20}$ & $0.85$&${}_{-\ 0.15}^{+\ 0.18}$ & $1.23$&${}_{-\ 0.22}^{+\ 0.27}$  \\
		$\mathcal{B}_{\tau\tau}/\mathcal{B}_{ZZ}$ & $0.76$&${}_{-\ 0.21}^{+\ 0.26}$ & $0.86$&${}_{-\ 0.22}^{+\ 0.26}$ & $1.07$&${}_{-\ 0.30}^{+\ 0.37}$  \\
		$\mathcal{B}_{bb}/\mathcal{B}_{ZZ}$  & $0.20$&${}_{-\ 0.12}^{+\ 0.21}$ & $0.93$&${}_{-\ 0.28}^{+\ 0.38}$  & $0.84$ &${}_{-\ 0.27}^{+\ 0.37}$ \\
		$\mathcal{B}_{\mu\mu}/\mathcal{B}_{ZZ}$ & $-$ & ${}$ &$-$& & $0.63$&${}_{-\ 1.21}^{+\ 1.24} $  \\
		\noalign{\smallskip}\hline
		\hline
	\end{tabular}
\end{table}

To test our point of view, we perform a fit to these measurements by minimizing a $\chi^2$ function, which is defined as
\begin{eqnarray}\label{eq:xihiggs}
\chi^2_{\rm Higgs} = \sum_{i=1}^{28} \sum_{j=1}^{28}[E_i -T_i]C_{ij}^{-1}[E_j -T_j],
\end{eqnarray}
where $E_i$ denotes experimentally measured $\sigma_{ggF}\cdot B_{ZZ} $, $\sigma_i / \sigma_{ggF}$ or $\mathcal{B}_i / \mathcal{B}_{ZZ}$ and $T_i$ is its theoretical expectation. $C$ is a $28\times28$ covariance matrix, which can be constructed by using the standard errors and corresponding correlations between these measurements obtained from the original articles published.

\begin{figure}[!]
	\centering{\includegraphics[width=3.0in]{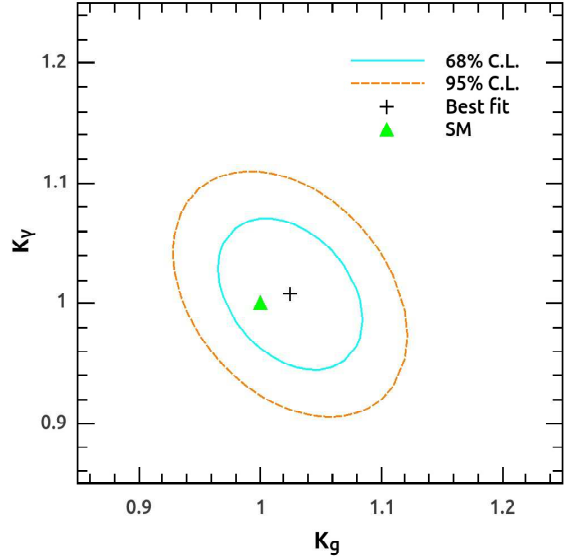}}
	\vspace*{8pt}
	\caption{Two dimensional likelihood contours at 68\% and 95\% C.L. in ($\kappa_g , \ \kappa_{\gamma}$) plane obtained from the LHC Run I and Run II Higgs data. The cross and triangle are the best-fit value and SM prediction, respectively. \protect\label{fig:kg-kr}}
\end{figure}

Assuming that BSM contributes to the loop processes only, we have
\begin{eqnarray}
\label{eq:gamh}
\kappa_h^2 = \frac{\Gamma_h}{\Gamma_h^{SM}}.
\end{eqnarray}
In this case, the coupling modifiers $\kappa_{\gamma}$ and $\kappa_g$ are free and other $\kappa_i$ are fixed to unity. The best fit to the measurements yields
\begin{eqnarray}
\kappa_{\gamma} = 1.008 \pm 0.042, \ \kappa_g = 1.025 \pm 0.040,
\end{eqnarray}
with the correlation between the two quantities $\rho=-0.34$.  Two dimensional likelihood contours  at 68\% and 95\% C.L. in ($\kappa_g , \ \kappa_{\gamma}$) plane are shown in Fig.~\ref{fig:kg-kr}. The fitting results are in good agreement within error with the SM predictions ( errors of $\kappa_{\gamma}$ and $\kappa_g$ are both reduced to about 4\% ), which further support our argument of NP only modifying loop-induced Higgs processes.

The appropriate cumulative distribution functions are used to obtain the upper bounds for this and following analysis, namely, 68\% (95\%) best-fit region satisfies $\chi^2 -\chi_{min}^2 \leq 0.99 \ (3.84)$ for one parameter, and $\chi^2 -\chi_{min}^2 \leq 2.28 \ (5.99)$ for two parameters.

\section{Scalar LQs}
\label{sec:3}

By using transformations under the SM gauge group $\mathcal{G}_{SM} = SU(3)_c\times SU(2)_L\times U(1)_Y$ as the classification criterion, there are six possible scalar LQ multiplets~\cite{Dorsner:2016wpm}: $S_{3}(\overline{\mathbf{3}},\mathbf{3},1/3), R_{2}(\mathbf{3},\mathbf{2},7/6), \widetilde{R}_{2}(\mathbf{3},\mathbf{2},1/6),\\\widetilde{S}_{1}(\overline{\mathbf{3}},\mathbf{1},4/3),S_{1}(\overline{\mathbf{3}},\mathbf{1},1/3), \bar{S}_{1}(\overline{\mathbf{3}}, \mathbf{1},-2/3)$. The first number, the second one and the last one within each brackets indicates the QCD representation, the weak isospin representation and the weak hypercharge, respectively.

The colorless vacuum requires that these colored scalars cannot acquire their masses via spontaneous symmetry breaking~\cite{Chang:2012ta}. Assuming weak components of single scalar LQ multiplet ($S$) to be degenerate at the electroweak scale, namely the mass of scalar LQ, $m_S$, is a free parameter, the Higgs portal interaction reads~\cite{Dorsner:2016wpm}
\begin{eqnarray}
\mathcal{L} \ni -\lambda_{S}(S_{ia}^{\dagger} S_{ia})(H_j^{\dagger}H_j)=-\lambda_{S} \upsilon (S_{ia}^{\dagger} S_{ia})h,
\label{eq:higgsportal}
\end{eqnarray}
where $i, j$ are weak indices, $a$ represents color index, $\lambda_{S}$ is the coupling constant for the LQ-Higgs-LQ vertex, $\upsilon$ is vacuum expectation of the Higgs boson with $\upsilon=246.22$ GeV.

Contributions of $S$ to loop-induced Higgs processes arise from Eq. (\ref{eq:higgsportal}), and are described by only two independent parameters, $\lambda_S$ and $m_S$. For convenience, a new parameter $\xi_{S}(\lambda_{S}, m_{S}^2) \equiv \lambda_{S}(\upsilon/m_{S})^2$ is introduced.

In the SM,  $W$ boson and top quark loops dominate the partial decay width of $ h \to \gamma \gamma$ decay. The partial decay width in presence of single scalar LQ $S$ is given by
\begin{eqnarray}
\label{eq:hrr}
\Gamma(h \to \gamma \gamma) = \frac{G_F \alpha_{em}^2 m_h^3}{128 \sqrt{2} \pi^3} \left| F_1(x_W) + \frac{4}{3} F_{1/2}(x_t) +  \sum_{i} \frac{\xi_{S}}{2} d(r_{S})Q_{S_i}^2 F_0(x_{S}) \right|^2,
\end{eqnarray}
where $G_{F}$ and $\alpha_{em}$ are the Fermi and fine-structure constants, respectively, $Q_{S_i}$ is electric charge of the weak component $S_i$ of single representation $S$, the sum of $i$ is taken over the weak components, $d(r_{S})$ represents the dimension of the color representation, and $x_i\equiv m_h^2/(4m_i^2)$ (i= W, t, $S$). The one-loop functions $F(x)$ read
\begin{eqnarray}
F_1(x) &=& \left[ x(2x+3) + 3(2x-1)f(x) \right]x^{-2}, \nonumber \\
F_{1/2}(x) &=& -2\left[ x + (x-1)f(x) \right]x^{-2}, \\
F_0(x) &=& \left[ x - f(x) \right]x^{-2}, \nonumber
\end{eqnarray}
with the function
\begin{eqnarray}
f({x})&=&\left\{ \begin{array}{cc}
\arcsin^{2}\sqrt{{x}} & {x} \leqslant 1\\
-\frac{1}{4}\left(\log\frac{1+\sqrt{1-{x^{-1}}}}{1-\sqrt{1-{x^{-1}}}}-i\pi\right)^{2} & {x}<1
\end{array}\right..
\end{eqnarray}
Then, one can obtain normalized modification of partial decay width of $h \to \gamma \gamma$ decay induced by single scalar LQ, which is expressed as~\cite{Dorsner:2016wpm}
\begin{eqnarray}
\label{kappa-hrr-S}
\frac{\Gamma_{h \to \gamma \gamma}}{\Gamma^{SM}_{h \to \gamma \gamma}} = |\kappa_{\gamma}|^2, \ {\rm where} \ \kappa_{\gamma} = 1-0.026 \xi_{S} d(r_S)\sum_{i}Q_{S_i}^2.
\end{eqnarray}

In the SM, top quark loop dominates the $ggF$ Higgs production cross section. In presence of single scalar LQ $S$, the leading order parton cross section of $ g g \to h$ at the partonic center mass of energy $\sqrt{\hat{s}}$ can be expressed as
\begin{eqnarray}
\hat{\sigma}_{LO}(g g \to h)= \frac{\sigma_0}{m_h^2}\delta(\hat{s}-m_h^2),
\end{eqnarray}
where $\sigma_0$ is proportional to the partial decay width of $h \to g g$ decay, which is given by
\begin{eqnarray}
\sigma_0 &=& \frac{8\pi^2}{m_h^3}\Gamma_{LO}(h \to g g) \nonumber \\
&=& \frac{G_F \alpha_s^2(\mu^2)}{512\sqrt{2}\pi} \left|F_{1/2}(x_t) + \sum_{i}^{N_{S_i}} \xi_{S} C(r_{S}) F_0(x_{S}) \right|^2,
\label{eq:ggh}
\end{eqnarray}
where $\alpha_s^2(\mu^2)$ represents the strong coupling constant, $F_0$ term induced by single scalar LQ $S$. $C(r_S)$ is the index of color representation of $S$ ( $C(r_S) = 1/2$ for color triplet ) and $N_{S_i}$ is the number of weak components of $S$. Effects of higher order QCD are neglected, since the ratio, $\sigma/(\sigma)_{SM}$, is found to be less sensitive to that~\cite{Gori:2013mia}. The normalized modification of $ggF$ Higgs production cross section induced by single scalar LQ is given by~\cite{Dorsner:2016wpm}
\begin{eqnarray}
\label{kappa-hgg-S}
\frac{\sigma_{gg \to h}}{\sigma^{SM}_{gg \to h}} = |\kappa_{g}|^2, \ {\rm where} \ \kappa_{g} = 1 + 0.24 \xi_{S} N_{S_i} C(r_S).
\end{eqnarray}

Thus for the case of single scalar LQ representation $S$ modifies the loop-induced Higgs processes, there is only one free parameter $\xi_S$ left.

To obtain $\xi_S$, we re-perform the Higgs fit by using $\xi_S$ to replace $\kappa_{\gamma}$ and $\kappa_g$ via Eqs. (\ref{kappa-hrr-S}) and (\ref{kappa-hgg-S}). Best values of $\xi_S$ with standard errors and 95\% C.L. intervals for all six scalar LQ representations are shown in Table~\ref{tab:fitreslut-slq}. Errors of $\xi_S$ for all scalar LQs obtained in this analysis are reduced more than half compared with previous analysis in Ref.~\cite{Dorsner:2016wpm}. But constraints
on $\xi_S$ for all scalar LQs are still too loose to acquire exact information for scalar LQs with TeV-scale masses. Table~\ref{SLQmasslimit} shows best values of scalar LQs masses and their lower limits at 95\% C.L. in the assumption of the portal coupling $\lambda_S = 1.0$. If LQs are insensitive to generation as well as their decay modes, the most stringent limits on the mass of scalar LQs reads $m_S > 1560$ GeV reported by the ATLAS collaboration~\cite{Aaboud:2019jcc}. Assuming $m_{S} = 1560$ GeV, best values of portal couplings $\lambda_S$ and their upper limits at 95\% C.L. obtained from Higgs fit are shown in Table~\ref{SLQcouplinglimit}.

The results are expected to be significantly improved at High Luminosity (HL)-LHC. Ref.~\cite{Cepeda:2019klc} reported the projections for Higgs couplings determinations at HL-LHC with an integrated luminosity of 3000 fb$^{-1}$. The precision on $\kappa_{\gamma}$ and $\kappa_g$ is expected to be 2.4\% and 3.1\% at the ATLAS experiment while that is 2.0\% and 2.5\% at the CMS experiment. Thus the precision on $\kappa_{\gamma}$ and $\kappa_g$ is expected to be 1.5\% and 1.9\% at HL-LHC by combining the ATLAS and CMS measurements of $\kappa_{\gamma}$ and $\kappa_g$. Then we can obtain the precision on $\xi_S$ expected at the HL-LHC via Eqs. (\ref{kappa-hrr-S}) and (\ref{kappa-hgg-S}). The approximate relation between errors of $\xi_S$ and $\kappa_{\gamma}$ and $\kappa_g$ read
\begin{eqnarray}
\delta_{\xi_S} \approx \left[ 0.24 N_{S_i} C(r_S) - 0.026 d(r_S)\sum_{i}Q_{S_i}^2 \right] \sqrt{\delta^2_{\kappa_g}+\delta^2_{\kappa_{\gamma}}}.
\end{eqnarray}
Compared to the present precision on $\xi_S$, the situation is expected to improve by a factor of 2.4 at the HL-LHC.

\begin{table}
	\caption{Constraints on LQs from the LHC Run I and II Higgs data for all scalar LQ representations, where $\xi_{S} = \lambda_S \upsilon^2 / m^2_S$.}
	\begin{center}
		\begin{tabular}{l|cc}
			\hline\hline
			\multirow{2}{*}{Scalar LQ} & \multicolumn{2}{c}{$\xi_{S} = \lambda_S \upsilon^2/m^2_S$}  \\
			\cline{2-3}
			& best fit  & 95\% C.L.  \\ \hline
			$S_3(\bar{\mathbf{3}}, \mathbf{3}, 1/3)$ &  0.060 $\pm$ 0.108  & [-0.173, 0.294]  \\
			$R_2(\mathbf{3}, \mathbf{2}, 7/6)$ &  0.032 $\pm$ 0.134  & [-0.241, 0.326]  \\
			$\tilde{R}_{2}(\mathbf{3}, \mathbf{2},1/6)$ &  0.115 $\pm$ 0.162  & [-0.237, 0.456]  \\
			$\tilde{S}_{1}(\bar{\mathbf{3}}, \mathbf{1},4/3)$ &  0.048 $\pm$ 0.245  & [-0.465, 0.604]  \\
			$S_{1}(\bar{\mathbf{3}},\mathbf{1},1/3)$ &  0.234 $\pm$ 0.316  & [-0.452, 0.895] \\
			$\bar{S}_{1}(\overline{\mathbf{3}}, \mathbf{1},-2/3)$ &  0.220 $\pm$ 0.329  & [-0.494, 0.917]  \\
			\hline
		\end{tabular}
	\end{center}
	\label{tab:fitreslut-slq}
\end{table}

\begin{table}
	\caption{ For the LQ-Higgs coupling $\lambda_S = 1.0$, best values and lower limits at 95\% C.L. of scalar LQs masses obtained from Higgs fit.}
	\begin{center}
		\begin{tabular}{l|cc}
			\hline\hline
			\multirow{2}{*}{Scalar LQ} & \multicolumn{2}{c}{$m_{S}$ ( $\lambda_S = 1.0$ )}  \\
			\cline{2-3}
			& best fit  & 95\% C.L.  \\ \hline
			$S_3(\bar{\mathbf{3}}, \mathbf{3}, 1/3)$ &  1005 GeV  & $>$ 454 GeV  \\
			$R_2(\mathbf{3}, \mathbf{2}, 7/6)$ &  1376 GeV  & $>$ 431 GeV  \\
			$\tilde{R}_{2}(\mathbf{3}, \mathbf{2},1/6)$ &  ~726 GeV  & $>$ ~364 GeV  \\
			$\tilde{S}_{1}(\bar{\mathbf{3}}, \mathbf{1},4/3)$ &  1124 GeV  & $>$ ~317 GeV  \\
			$S_{1}(\bar{\mathbf{3}},\mathbf{1},1/3)$ &  ~509 GeV  & $>$ ~260 GeV \\
			$\bar{S}_{1}(\overline{\mathbf{3}}, \mathbf{1},-2/3)$ &  ~525 GeV  & $>$ ~257 GeV  \\
			\hline
		\end{tabular}
	\end{center}
	\label{SLQmasslimit}
\end{table}

\begin{table}
	\caption{ For $m_{S} = 1000$ GeV, best values and upper limits at 95\% C.L. of the size of LQ-Higgs coupling $|\lambda_S|$ for scalar LQs obtained from Higgs fit.}
	\begin{center}
		\begin{tabular}{l|cc}
			\hline\hline
			\multirow{2}{*}{Scalar LQ} & \multicolumn{2}{c}{$\lambda_S$ ( $m_{S} = 1560$ GeV )}  \\
			\cline{2-3}
			& \quad  best fit \quad \ & 95\% C.L.   \\ \hline
			$S_3(\bar{\mathbf{3}}, \mathbf{3}, 1/3)$ &  ~2.4  & $<$ 11.8   \\
			$R_2(\mathbf{3}, \mathbf{2}, 7/6)$ &  ~1.3  & $<$ 13.1   \\
			$\tilde{R}_{2}(\mathbf{3}, \mathbf{2},1/6)$ &  ~4.6  & $<$ 18.3  \\
			$\tilde{S}_{1}(\bar{\mathbf{3}}, \mathbf{1},4/3)$ &  1.9  & $<$ 24.2  \\
			$S_{1}(\bar{\mathbf{3}},\mathbf{1},1/3)$ &  ~9.4  & $<$ 35.9   \\
			$\bar{S}_{1}(\overline{\mathbf{3}}, \mathbf{1},-2/3)$ &  ~8.8  & $<$ 36.8  \\
			\hline
		\end{tabular}
	\end{center}
	\label{SLQcouplinglimit}
\end{table}

\section{Vector LQ $U_1(\mathbf{3},\mathbf{1},2/3)$ in the '4321' model}
\label{sec:4}

Now we consider contributions of vector LQ $U_1(\mathbf{3},\mathbf{1},2/3)$ to the loop-induced Higgs processes $gg \to h$ and $h \to \gamma \gamma$, which LQ performs quite well in explaining both anomalies of $R_{D^{(*)}}$ and $R_{K^{(*)}}$. Our study in framework of the '4321' model~\cite{DiLuzio:2017vat,DiLuzio:2018zxy}. We first briefly review the '4321' model, then we study contributions of $U_1$ to the loop-induced Higgs processes. Constraints on the interactions of $U_1$ with the Higgs boson from LHC Higgs data is obtained. Further more, we obtain constraints on the VEVs $\upsilon_3$ and $\upsilon_1$ of new scalar fields $\Omega_3$ and $\Omega_1$ in the model.

\subsection{The '4321' model}

The model gauge group is expressed as $\mathcal{G}_{4321} = SU(4)\times SU(3)'\times SU(2)_L\times U(1)'$, for which $H^{\alpha}_{\mu}, G'^a_{\mu}, W^i_{\mu}, \\ B'_{\mu}$ denote corresponding gauge fields, $g_4, g_3, g_2, g_1$ the gauge couplings and $T^{\alpha}, T^{a}, T^i, Y'$ the generators, where the indices $\alpha = 1,...,15,\ a= 1,...,8,\ i=1,...,3$. The generators are normalized in such a way that $\text{Tr} T^A T^B = \frac{1}{2}\delta^{AB}$. The SM gauge symmetry $SU(3)_c\times U(1)_Y$ is embedded in $SU(4)\times SU(3)'\times U(1)'$.

The model comprises four scalar representations: $\Omega_3(\bar{\mathbf{4}},\mathbf{3},\mathbf{1},1/6)$, $\Omega_1(\bar{\mathbf{4}},\mathbf{1},\mathbf{1},-1/2)$, $\Omega_{15}(\overline{\mathbf{15}},\mathbf{1},\mathbf{1},0)$ and $\Phi(\mathbf{1},\mathbf{1},\mathbf{2},1/2)$, where $\Omega_3$ and $\Omega_1$ are respectively a $4\times3$ matrix and a $4$-vector transforming as $\Omega_3 \to U^*_4 \Omega_3 U^T_{3'}$ and $\Omega_1 \to U^*_4 \Omega_1$ under $SU(4)\times SU(3)'$ and $H$ is the Higgs doublet ( in this analysis we neglect the effect of $\Omega_{15}$ ). Phenomenological considerations suggest : $ \langle \Omega_3 \rangle > \langle \Omega_1 \rangle > \langle \Phi \rangle $. According to Ref.~\cite{DiLuzio:2018zxy}, the most general scalar potential involving $\Omega_{3,1}$ and $H$ can be written as
\begin{eqnarray}
\label{scalar-potential}
V = &+& \mu_3^2{\rm Tr}(\Omega_{3}^{\dagger}\Omega_{3}) +
\lambda_1 \left({\rm Tr}(\Omega_{3}^{\dagger}\Omega_{3}) - \frac{3}{2}\upsilon_3^2 \right)^2 + \lambda_2
{\rm Tr} \left(\Omega_{3}^{\dagger}\Omega_{3} - \frac{1}{2}\upsilon_3^2 \right)^2 \nonumber \\&+&\mu_1^2|\Omega_{1}|^2 + \lambda_3 \left( |\Omega_{1}|^2 - \frac{1}{2}\upsilon_1^2 \right)^2 + \lambda_4 \left({\rm Tr}(\Omega_{3}^{\dagger}\Omega_{3}) - \frac{3}{2}\upsilon_3^2 \right) \left( |\Omega_{1}|^2 - \frac{1}{2}\upsilon_1^2 \right) \nonumber \\
&+& \lambda_5 \Omega_{1}^{\dagger}\Omega_{3}\Omega_{3}^{\dagger}\Omega_{1} + \lambda_6 \left( [\Omega_{3} \Omega_{3} \Omega_{3} \Omega_{1}]_1 + {\rm h.c.} \right) + \mu^2_{\Phi}\Phi^{\dagger}\Phi + \lambda_7 \left(\Phi^{\dagger}\Phi - \frac{\upsilon^2}{2} \right)^2 \nonumber \\
&+& \lambda_8 \left({\rm Tr}(\Omega_{3}^{\dagger}\Omega_{3}) - \frac{3}{2}\upsilon_3^2 \right) \left(\Phi^{\dagger}\Phi - \frac{\upsilon^2}{2} \right) + \lambda_9 \left( |\Omega_{1}|^2 - \frac{1}{2}\upsilon_1^2 \right) \left(\Phi^{\dagger}\Phi - \frac{\upsilon^2}{2} \right) .
\end{eqnarray}
where $\left[ \Omega_3 \Omega_3 \Omega_3 \Omega_1 \right]_1 \equiv \epsilon_{\alpha \beta \gamma \delta} \epsilon^{a b c} \left( \Omega_3 \right)^{\alpha}_a \left( \Omega_3 \right)^{\beta}_b \left( \Omega_3 \right)^{\gamma}_c \left( \Omega_1 \right)^{\delta}$.
VEV configurations~\cite{DiLuzio:2018zxy}
\begin{eqnarray}
\langle \Omega_3 \rangle = \frac{1}{\sqrt{2}} \begin{pmatrix}
\upsilon_3 & 0 & 0 \\
0 & \upsilon_3 & 0 \\
0 & 0 & \upsilon_3 \\
0 & 0 & 0
\end{pmatrix} \ , \ \langle \Omega_1 \rangle = \frac{1}{\sqrt{2}} \begin{pmatrix}
0\\
0\\
0\\
\upsilon_1
\end{pmatrix},
\end{eqnarray}
together with $\mu^2_3 = -3\lambda_6 \upsilon_3 \upsilon_1$, $\mu^2_1 = - 3\lambda_6 \upsilon^2_3 / \upsilon_1$ and $\mu^2_h = 0$ in Eq.~(\ref{scalar-potential}) ensure the proper $\mathcal{G}_{4321} \to \mathcal{G}_{SM}$ breaking. Under $\mathcal{G}_{SM}$, $\Omega_3$ and $\Omega_1$  decomposed as: $\Omega_3 \to \mathbb{S}_3(\mathbf{1},\mathbf{1},0)\oplus \mathbb{T}_3(\mathbf{3},\mathbf{1},2/3)\oplus \mathbb{O}_3(\mathbf{8},\mathbf{1},0)$ and $\Omega_1 \to \mathbb{S}_1(\mathbf{1},\mathbf{1},0)\oplus \mathbb{T}_1^*(\mathbf{3},\mathbf{1},2/3)$. The final breaking of $\mathcal{G}_{SM}$ proceeds via the Higgs doublet field acquiring a VEV $\langle \Phi \rangle = (0 \ \upsilon)^T/\sqrt{2}$, with $\upsilon = 246.22$ GeV.

The covariant derivatives of $\Omega_3$, $\Omega_1$ and $\Phi$ are given by
\begin{eqnarray}
D_{\mu} \Omega_3 &=&  \partial_{\mu} \Omega_3 + {\rm i}g_4 H^{\alpha}_{\mu} T^{*\alpha} \Omega_3 - {\rm i}g_3 G^{'a}_{\mu} T^{a} \Omega_3 - \frac{1}{6}{\rm i} g_1 B'_{\mu} \Omega_3, \nonumber \\
D_{\mu} \Omega_1 &=&  \partial_{\mu} \Omega_1 + {\rm i}g_4 H^{\alpha}_{\mu} T^{*\alpha} \Omega_1  + \frac{1}{2} {\rm i} g_1 B'_{\mu} \Omega_1, \nonumber \\
D_{\mu} \Phi_{~} &=&  \partial_{\mu} \Phi_{~} - {\rm i}g_2 W^{i}_{\mu} T^{i} \Phi  - \frac{1}{2} {\rm i} g_1 B'_{\mu} \Phi \label{eq:CDS}
\end{eqnarray}

In the model, the mass of $U_1$ and corresponding mass eigenstate expressed in terms of the original gauge fields are given by~\cite{DiLuzio:2018zxy}
\begin{eqnarray}
m_U = \dfrac{1}{2} g_4 \sqrt{\upsilon_3^2 + \upsilon_1^2}, \label{eq:mU}
\end{eqnarray}
and
\begin{eqnarray}
U_{1\mu}^{1,2,3} = \frac{1}{2} \left( H_{\mu}^{9,11,13} - {\rm i}H_{\mu}^{10,12,14} \right).
\end{eqnarray}
Then we obtain Feynman rules of $U_1$ interactions to scalars
\begin{eqnarray}
\begin{pmatrix}U^{~}_{1 \mu} \\ U^{*}_{1 \nu}\\ \mathbb{S}^{(*)}_{3}\end{pmatrix} &:& ~\dfrac{\rm i}{2} g^2_4 \frac{\upsilon_{3}}{4\sqrt{3}} g_{\mu \nu} ,\qquad  \begin{pmatrix}U^{~}_{1 \mu} \\ U^{*}_{1 \nu}\\ \mathbb{S}^{(*)}_{1}\end{pmatrix} : ~\dfrac{\rm i}{2} g^2_4 \frac{\upsilon_{1}}{2\sqrt{2}} g_{\mu \nu} , \qquad \qquad \quad \label{V-uus}
\end{eqnarray}
From Eq.~(\ref{eq:CDS}) we can see that $U_1$ can not couple to the Higgs doublet $\Phi$ directly. $U_1$ interacts with the Higgs boson $h$ via the mixing of $\phi^{(*)}$ and representations $\mathbb{S}^{(*)}_{3,1}(\mathbf{1},\mathbf{1},0)$ after the final SM breaking, where $\phi$ represents the neutral component of the Higgs doublet and $\mathbb{S}_{3,1}(\mathbf{1},\mathbf{1},0)$ are decompositions of $\Omega_{3,1}$ under the SM symmetry. In the basis ($\mathbb{S}_3, \mathbb{S}^*_3, \mathbb{S}_1, \mathbb{S}^*_1, \phi,\phi^*$), singlet spectrum are expressed as
\vspace{1mm}
\begin{normalsize}
	\begin{multline}
	\mathcal{M}^2_{S} = \\
	\left(
	\begin{array}{cccccc}
	\mathcal{M}^2_1 &
	\mathcal{M}^2_2 &
	\mathcal{M}^2_3 &
	\frac{1}{2} \sqrt{\frac{3}{2}} \lambda_4 v_1 v_3 & \frac{1}{2} \sqrt{\frac{3}{2}} \lambda_8 v v_3 & \frac{1}{2} \sqrt{\frac{3}{2}} \lambda_8 v v_3  \\
	\mathcal{M}^2_2 &
	\mathcal{M}^2_1 &
	\frac{1}{2} \sqrt{\frac{3}{2}} \lambda_4 v_1 v_3 &
	\mathcal{M}^2_3 & \frac{1}{2} \sqrt{\frac{3}{2}} \lambda_8 v v_3 & \frac{1}{2} \sqrt{\frac{3}{2}} \lambda_8 v v_3 \\
	\mathcal{M}^2_3 &
	\frac{1}{2} \sqrt{\frac{3}{2}} \lambda_4 v_1 v_3 &
	\lambda_3 v_1^2 &
	\lambda_3 v_1^2-3\lambda_6\frac{v_3^3}{v_1} & \frac{1}{2} \lambda_9 v v_1 & \frac{1}{2} \lambda_9 v v_1 \\
	\frac{1}{2} \sqrt{\frac{3}{2}} \lambda_4 v_1 v_3 &
	\mathcal{M}^2_3 &
	\lambda_3 v_1^2-3\lambda_6\frac{v_3^3}{v_1} &
	\lambda_3 v_1^2 & \frac{1}{2} \lambda_9 v v_1 & \frac{1}{2} \lambda_9 v v_1 \\
	\frac{1}{2} \sqrt{\frac{3}{2}} \lambda_8 v v_3 & \frac{1}{2} \sqrt{\frac{3}{2}} \lambda_8 v v_3 &  \frac{1}{2} \lambda_9 v v_1 & \frac{1}{2} \lambda_9 v v_1 & \lambda_7 \upsilon^2 & \lambda_7 \upsilon^2 \\
	\frac{1}{2} \sqrt{\frac{3}{2}} \lambda_8 v v_3 & \frac{1}{2} \sqrt{\frac{3}{2}} \lambda_8 v v_3 &  \frac{1}{2} \lambda_9 v v_1 & \frac{1}{2} \lambda_9 v v_1 & \lambda_7 \upsilon^2 & \lambda_7 \upsilon^2
	\end{array}
	\right),
	\end{multline}
\end{normalsize}
where
\begin{eqnarray}
\mathcal{M}^2_1 = \frac{1}{2} \left( 3 \lambda_1 + \lambda_2 \right) v_3^2+3 \lambda_6 v_1 v_3, \ \mathcal{M}^2_2 = \frac{1}{2} \left( 3 \lambda_1 + \lambda_2 \right) v_3^2-\frac{3}{2} \lambda_6 v_1 v_3, \ \mathcal{M}^2_3 = \sqrt{\frac{3}{2}} \left( 3 \lambda_6 v_3^2 + \frac{1}{2} \lambda_4 v_1 v_3 \right). \nonumber
\end{eqnarray}
It turns out  $\mathcal{M}^2_{S} = 4$. Two massless modes correspond to eigenvectors
\begin{eqnarray}
S^Z_{GB} = \dfrac{1}{\sqrt{2}} \left( 0, 0, 0, 0, -1, 1 \right),
\end{eqnarray}
and
\begin{eqnarray}
S^{Z'}_{GB} = \dfrac{1}{\sqrt{\frac{2}{3}\upsilon^2_3 + \upsilon^2_1}} \left( \dfrac{\upsilon_3}{\sqrt{3}}, \ -\dfrac{\upsilon_3}{\sqrt{3}}, \ -\dfrac{\upsilon_1}{\sqrt{2}}, \ \dfrac{\upsilon_1}{\sqrt{2}}, \ 0, \ 0 \right),
\end{eqnarray}
which are associated to the longitudinal degrees of freedom of the $Z$ and $Z'$, respectively. One of four non-zero eigenvalues as well as corresponding eigenvector can be also easily obtained as
\begin{eqnarray}
\label{eq:ms0}
\mathcal{M}^2_{S_0} = \dfrac{3\lambda_6 \upsilon_3 \left( 2\upsilon^2_3 + 3\upsilon^2_1 \right)}{2 \upsilon_1},
\end{eqnarray}
and
\begin{eqnarray}
S_0 =\dfrac{1}{\sqrt{\frac{2}{3}\upsilon^2_3 + \upsilon^2_1}} \left( -\dfrac{\upsilon_1}{\sqrt{2}}, \ \dfrac{\upsilon_1}{\sqrt{2}}, \ -\dfrac{\upsilon_3}{\sqrt{3}}, \ \dfrac{\upsilon_3}{\sqrt{3}}, \ 0, \ 0 \right).
\end{eqnarray}

Precisely acquiring mass eigenvalue of the would-be Higgs boson and corresponding eigenvector is difficult, unless conditions such as precise value of the Higgs mass obtained from experiment as well as other constraints are applied. We assume the Higgs boson with mass value of $125.09$ GeV corresponds to normalized eigenvector
\begin{eqnarray}
\label{eq:heigenvector}
h = \left( \lambda_{\mathbb{S}_3}, \ \lambda_{\mathbb{S}^*_3}, \ \lambda_{\mathbb{S}_1}, \ \lambda_{\mathbb{S}^*_1}, \ \lambda_{\phi}, \ \lambda_{\phi^*} \right),
\end{eqnarray}
where $\lambda_{\mathbb{S}^{(*)}_{3}}$, $\lambda_{\mathbb{S}^{(*)}_{1}}$ and $\lambda_{\phi^{(*)}}$ represent mixing constants with $\lambda_{\mathbb{S}_{3}} = \lambda_{\mathbb{S}^{*}_{3}}$, $\lambda_{\mathbb{S}_{1}} = \lambda_{\mathbb{S}^{*}_{1}}$ and $\lambda_{\phi} = \lambda_{\phi^{*}}$, since $h$ is a real field. According to Eq.~(\ref{V-uus}), Feynman rule of $U_{1 \mu} U^{*}_{1 \nu}h$ should be expressed as
\begin{eqnarray}
\label{eq:uuh1}
U_{1 \mu} U^{*}_{1 \nu}h \ : \quad \frac{\rm i}{2} g^2_4 \left( \frac{\upsilon_{3}}{2\sqrt{3}} \lambda_{\mathbb{S}_3} + \frac{\upsilon_{1}}{\sqrt{2}} \lambda_{\mathbb{S}_1} \right) g_{\mu \nu}.
\end{eqnarray}
We do not intend to further solve these mixing parameters $\lambda_{\mathbb{S}^{(*)}_{3,1}}$. For convenience, we re-express the Feynman rule as
\begin{eqnarray}
U_{1 \mu} U^{*}_{1 \nu}h \ : \quad \frac{\rm i}{2} g^2_4 \upsilon g_{\mu \nu} \dfrac{\upsilon_3}{\upsilon} \lambda_V, \label{V-uuh}
\end{eqnarray}
where the $U_1$-Higgs coupling $\lambda_V = \frac{\lambda_{\mathbb{S}_3}}{2\sqrt{3}} + \frac{\upsilon_1}{\upsilon_3}\frac{\lambda_{\mathbb{S}_1}}{\sqrt{2}}$, which is expected to be small according to current Higgs measurements analyses in Section~\ref{sec:2}.

We now consider interactions among gauge bosons. The interactions are obtained from the gauge kinetic term~\cite{DiLuzio:2018zxy}
\begin{eqnarray}
\mathcal{L}_{gauge} = -\dfrac{1}{4} H^{\alpha}_{\mu \nu} H^{\alpha,\mu \nu} -\dfrac{1}{4} G'^{a}_{\mu \nu} G'^{a,\mu \nu} -\dfrac{1}{4} W^{i}_{\mu \nu} W^{i,\mu \nu} -\dfrac{1}{4} B'_{\mu \nu} B'^{\mu \nu},
\end{eqnarray}
where definitions of field strengths $H^{\alpha}_{\mu \nu}$, $G'^{a}_{\mu \nu}$, $W^{i}_{\mu \nu}$ and $B'_{\mu \nu}$ see ~\cite{DiLuzio:2018zxy}.

Prior to electroweak symmetry breaking, the massless $SU(3)_c\times U(1)_Y$ degrees of freedom of $\mathcal{G}_{SM}$ expressed in terms of the original gauge fields are  given by~\cite{DiLuzio:2018zxy}
\begin{eqnarray}
&g_{\mu}^a = \dfrac{g_3 H_{\mu}^a + g_4 G'^a_{\mu}}{\sqrt{g_4^2 + g_3^2}} , \\ &B_{\mu} = \dfrac{\sqrt{\frac{2}{3}}g_1 H_{\mu}^{15} + g_4 B'_{\mu}}{\sqrt{g_4^2 + \frac{2}{3}g_1^2}}.
\end{eqnarray}
The SM gauge couplings are matched as~\cite{DiLuzio:2018zxy}
\begin{eqnarray}
&g_s = \dfrac{g_4 g_3}{\sqrt{g_4^2 + g_3^2}} , \label{eq:gsg4g3} \\
&g_Y = \dfrac{g_4 g_1}{\sqrt{g_4^2 + \frac{2}{3}g_1^2}}. \label{eq:gsg4g1}
\end{eqnarray}

Then, one can obtain Feynman rules related to $U_1$ interactions to the SM gauge boson $\gamma$ and $g$,
\begin{eqnarray}
\begin{pmatrix}
U^{~}_{1 \mu}(k_1) \\ U^{*}_{1 \nu}(k_2) \\ A_{\rho}(k_3)
\end{pmatrix} &:& -{\rm i}  \dfrac{\frac{2}{3}g_4g_1 {\rm cos}(\theta_W)}{\sqrt{g_4^2 + \frac{2}{3}g_1^2}} V_{\mu \nu \rho} \left(k_1, k_2, k_3\right)  = - {\rm i} e Q_{U}V_{\mu \nu \rho} \left(k_1, k_2, k_3\right), \label{V-uua}
\end{eqnarray}
\begin{eqnarray}
\begin{pmatrix}U^{~}_{1 \mu}(k_1) \\ U^{*}_{1 \nu}(k_2) \\ A_{\rho}(k_3) \\ A_{\sigma}(k_4) \end{pmatrix} &:& ~
{\rm i} (e Q_{U})^2 \left( g_{\mu \rho}g_{\nu \sigma} + g_{\mu \sigma} g_{\nu \rho} - 2g_{\mu \nu}g_{\rho \sigma}\right), \qquad \qquad \qquad \qquad  \label{V-uuaa}
\end{eqnarray}
\begin{eqnarray}
\begin{pmatrix}U^i_{1 \mu}(k_1) \\ U^{*j}_{1 \nu}(k_2) \\ g^a_{\rho}(k_3) \end{pmatrix} &:&
- {\rm i}  \dfrac{g_4g_3 }{\sqrt{g_4^2 + g_3^2}}T^a_{ij} V_{\mu \nu \rho} \left(k_1, k_2, k_3\right) = - {\rm i} g_s T^a_{ij} V_{\mu \nu \rho} \left(k_1, k_2, k_3\right),  \label{V-uug}
\end{eqnarray}
\begin{eqnarray}
\begin{pmatrix}U^i_{1 \mu}(k_1) \\ U^{*j}_{1 \nu}(k_2) \\ g^a_{\rho}(k_3) \\ g^b_{\sigma}(k_4)\end{pmatrix} &:& ~{\rm i} g^2_s \delta_{ij} \dfrac{\delta_{ab}}{4} \left(g_{\mu \rho}g_{\nu \sigma} + g_{\mu \sigma} g_{\nu \rho} - 2g_{\mu \nu}g_{\rho \sigma} \right), \qquad \qquad \qquad \qquad  \label{V-uugg}
\end{eqnarray}
where $\theta_W$ is the Weinberg angle, the function $V_{\mu \nu \rho} \left(k_1, k_2, k_3\right)$ is defined as
\begin{eqnarray}
V_{\mu \nu \rho} \left(k_1, k_2, k_3\right)
= g_{\mu \nu} (k_2 - k_1)_{\rho} + g_{\nu \rho}(k_3 - k_2)_{\mu} - g_{\rho \mu}(k_1 - k_3)_{\nu}, \nonumber
\end{eqnarray}
with $k_i$ being four-momentum of the $i$-th particle ( direction towards the vertex is specified to be positive ).

\subsection{Constraints on $U_1$ from Higgs data}

By using Eqs.~(27,33-36), we obtain the partial decay width of $ h \to \gamma \gamma$ and cross section of $ g g \to h$ in presence of $U_1$
\begin{eqnarray}
\label{eq:hrr-U}
\Gamma(h \to \gamma \gamma) = \frac{G_F \alpha_{em}^2 m_h^3}{128 \sqrt{2} \pi^3} \left| F_1(x_W) + \frac{4}{3} F_{1/2}(x_t) + \xi_V d(r_{U})Q_{U}^2 F_1(x_{U}) \right|^2 ,
\end{eqnarray}
and
\begin{eqnarray}
\sigma_0 = \frac{G_F \alpha_s^2(\mu^2)}{512\sqrt{2}\pi} \left|F_{1/2}(x_t) + \xi_V F_1(x_{U}) \right|^2,
\label{eq:ggh-U}
\end{eqnarray}
where
\begin{eqnarray}
\xi_V = \dfrac{g^2_4 \lambda_V \upsilon \upsilon_3}{4 m^2_U} = \dfrac{\lambda_V \upsilon \upsilon_3}{\upsilon^2_3 + \upsilon^2_1} .
\label{eq:Xi_V}
\end{eqnarray}
In obtaining Eq.~(\ref{eq:Xi_V}), we have used mass expression Eq.~(\ref{eq:mU}). Eq.~(\ref{eq:Xi_V}) shows that the '4321' model's $U_1$ modifications to the loop-induced Higgs processes depend on $U_1$-Higgs coupling $\lambda_V$ and new VEVs $\upsilon_3$ and $\upsilon_1$ in the model rather than the mass of $U_1$ and gauge coupling $g_4$. This means that once $\xi_V$ is determined from the Higgs fit one can determine $\upsilon_3$ and $\upsilon_1$ by using $\xi_V$ together with other condition such as the mass of $U_1$ determined from colliders.

For single vector LQ $U_1$ modifying partial decay width of $ h \to \gamma \gamma$ and cross section of $ g g \to h$, coupling modifiers $\kappa_{\gamma}$ and $\kappa_g$ are expressed with $\xi_V$, which read
\begin{eqnarray}
\label{kappa-rr-gg-v}
\kappa_{\gamma} = 1+1.44 \xi_V \ {\rm and} \ \kappa_g = 1 - 5.09 \xi_V.
\end{eqnarray}

To obtain the size of $U_1$ interaction with the Higgs boson, we re-analyze the Higgs data by using Eq.~(\ref{kappa-rr-gg-v}). The best value with standard error and 95\% C.L. intervals of $\xi_V$ obtained from the Higgs fit are
\begin{eqnarray}
\label{limit-V}
\xi_V = -0.005 \pm 0.008 ,
\end{eqnarray}
and
\begin{eqnarray}
\label{limit-V95}
\xi_V \in \left[ -0.021, \ 0.011 \right].
\end{eqnarray}

\begin{figure}
	\centering
	\subfigure[]{
		\label{fig:subfig:a} %% label for first subfigure
		\includegraphics[width=3.0in]{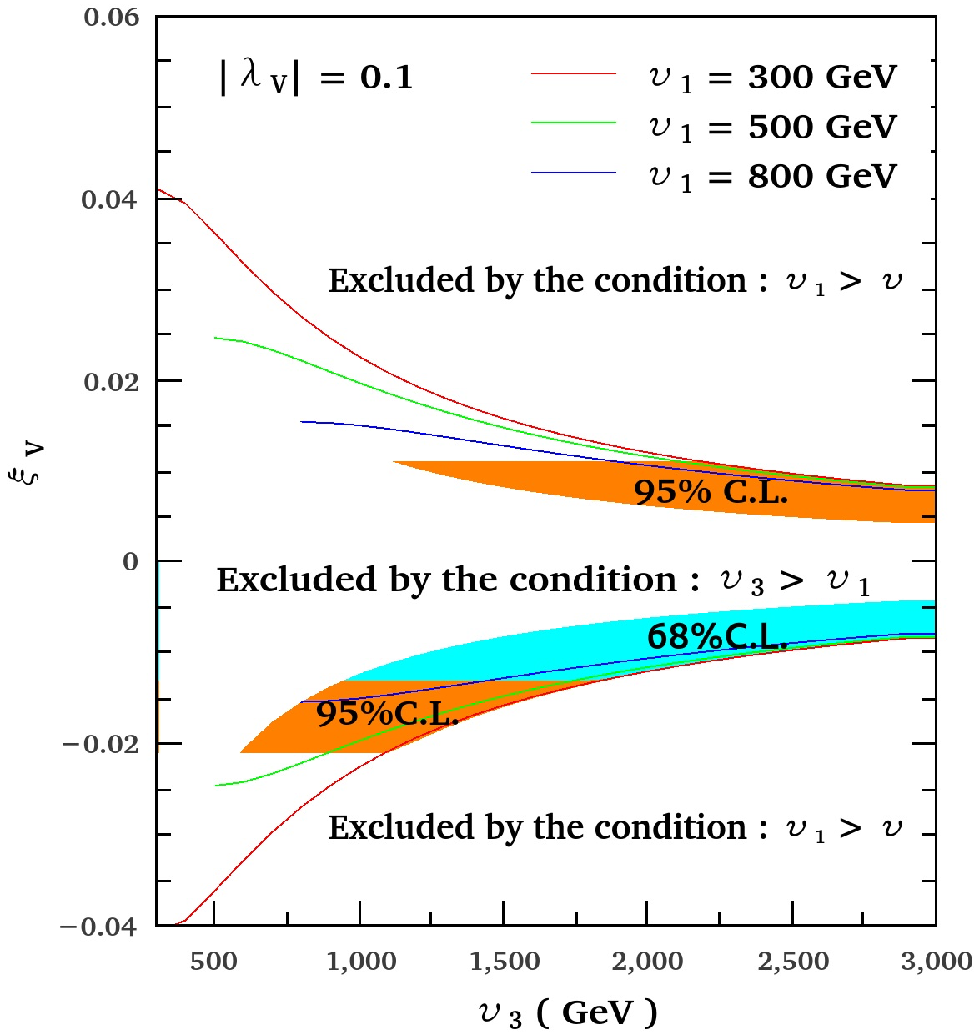}}
	\hspace{0.2in}
	\subfigure[]{
		\label{fig:subfig:b} %% label for second subfigure
		\includegraphics[width=3.0in]{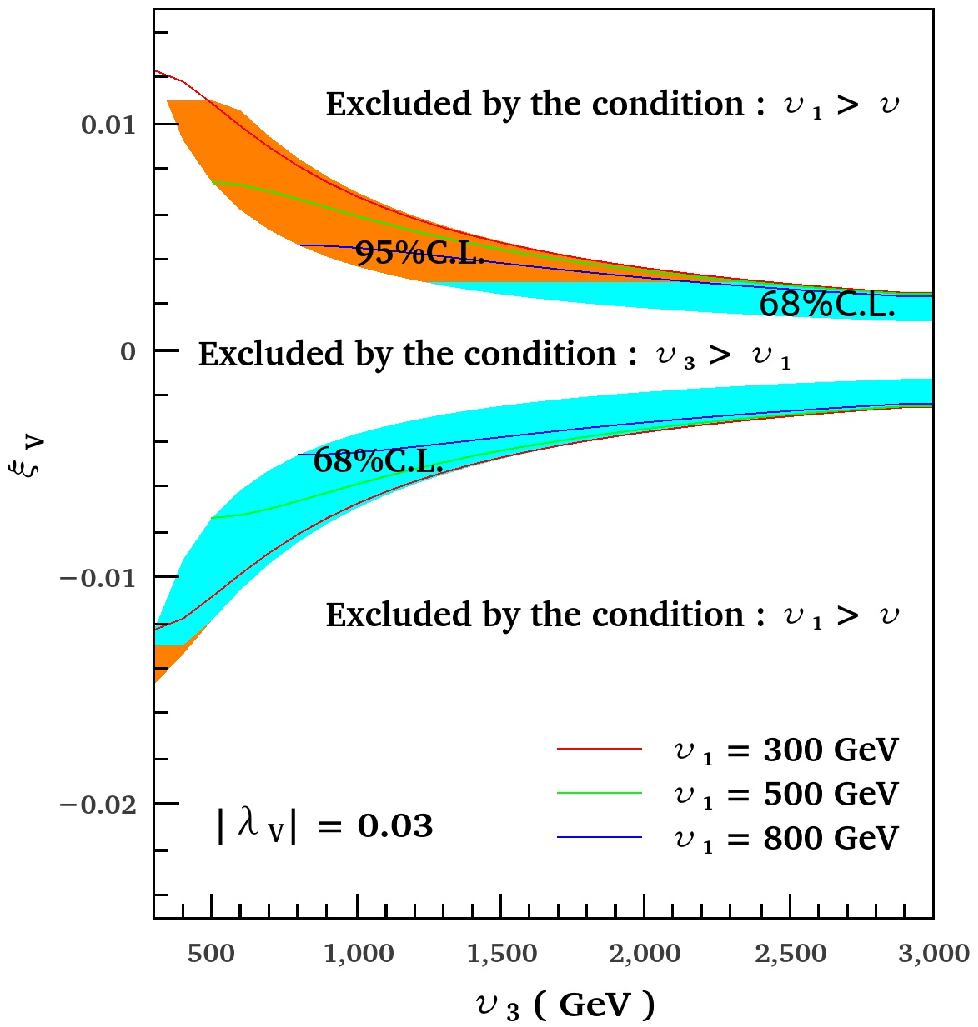}}
	\caption{ Limits on $\upsilon_3$ and $\upsilon_1$ from combination of phenomenological condition, $\upsilon_3 > \upsilon_1> \upsilon$, and constraints from Higgs fit. The cyan region is allowed at 68\% C.L. from the Higgs fit while the orange area is allowed at 95\% C.L.. The coupling size is assumed with value of (a) $|\lambda_V| = 0.1$ and (b) $|\lambda_V| = 0.03$. The red ( green, orange ) line represents $\xi_V$ varying as a function of $\upsilon_3$, in assuming $\upsilon_1$ with fixed value of 300 ( 500, 800 ) GeV.  \protect\label{fig:v3-XiV}}
\end{figure}

For $U_1$-Higgs coupling with value of one-third (-tenth) of the electromagnetic coupling strength, $|\lambda_V| = 0.1 (0.03)$, $\xi_V$ varying as a function of $\upsilon_3$ for a fixed value of $\upsilon_1$  and combined limits on $\upsilon_3$ and $\upsilon_1$ from the condition, $\upsilon_3 > \upsilon_1 > \upsilon$, as well as current Higgs data, are shown in Fig.~\ref{fig:v3-XiV}. From Fig.~\ref{fig:v3-XiV} one can see that we still need more precise Higgs measurements, since at least the sign of $\lambda_V$ has not been determined yet from current Higgs data. It should be noted that the result of precision on $\xi_S$ is also applicable to $\xi_V$, which means the precision on $\xi_V$ is expected to improve by a factor of 2.4 compared with present situation at HL-LHC.

\subsection{Combined limits on the relation of $\lambda_V$ and $\upsilon_{3,1}$}

We can further constrain the relation of $\lambda_V$ and $\upsilon_{3,1}$ by combined limits on $\xi_V$ obtained in this analysis and $m_U$ obtained from direct searches at colliders as well as measurements of $R_{D^{(*)}}$ and $R_{K^{(*)}}$. Details of obtaining the constraints on $U_1$ from $B$-anomalies are shown in appendix \ref{sec:A}

\begin{figure}[!]
	\centering{\includegraphics[width=3.0in]{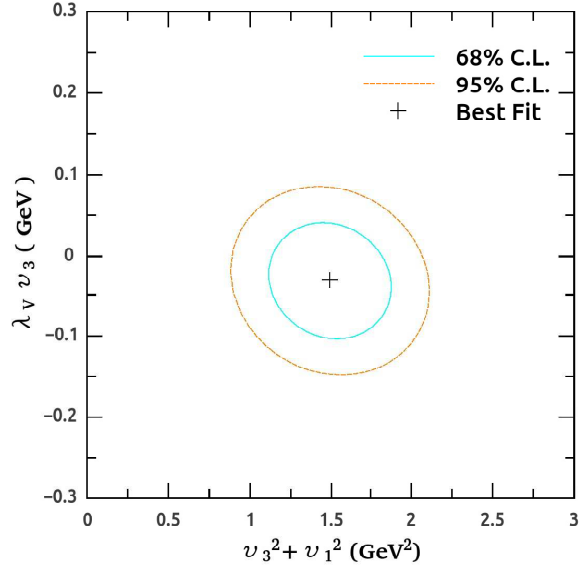}}
	\vspace*{8pt}
	\caption{Two dimensional likelihood contours at 68\% and 95\% C.L. in ($\upsilon^2_3 + \upsilon^2_1 , \ \lambda_V \upsilon_3$) plane obtained from combined limits of the LHC Higgs data together with constraints from $B$-anomalies measurements. The cross is the best-fit value. \protect\label{fig:vvlv}}
\end{figure}

Current lower limits on masses of vector LQs with decay mode $LQ \to t \nu / b \tau$ is $m_{LQ} > 1530$ GeV reported by the CMS collaboration~\cite{Sirunyan:2018kzh}. For $g_4 = 3.5$,
\begin{eqnarray}
\upsilon^2_3 +\upsilon^2_1 = \frac{4}{g^2_4} m^2_U > (874 \ {\rm GeV})^2,
\end{eqnarray}
which is looser than the constraints from $B$-anomalies. Thus we consider combined constraints from the LHC Higgs data and $B$-anomalies measurements, which is performed via minimizing
\begin{eqnarray}
\chi^2 = \chi^2_{\rm Higgs} + \chi^2_B,
\end{eqnarray}
where $\chi^2_{\rm Higgs}$ has been shown in Eq.~(\ref{eq:xihiggs}) and $\chi^2_B$ is explained in Eq.~(\ref{eq:xiB}). Assuming tree level contributions induced by $U_1$ dominant the NP contributions to $B$-anomalies. Fig.~{\ref{fig:vvlv}} shows two dimensional likelihood contours at 68\% and 95\% C.L. in ($\upsilon^2_3 + \upsilon^2_1 , \ \lambda_V \upsilon_3$) plane obtained from combination of the LHC Higgs data together with measurements of $B$-anomalies. Best values of $\upsilon^2_3 + \upsilon^2_1$ and $\lambda_V \upsilon_3$ read
\begin{eqnarray}
\upsilon^2_3 + \upsilon^2_1 = 1.496 \pm 0.250 \ {\rm TeV^2} \ {\rm and} \  \lambda_V \upsilon_3 = -0.0315 \pm 0.0473 \ {\rm TeV}.
\end{eqnarray}

\begin{figure}
	\centering
	\subfigure[]{
		\label{fig:subfig:a} %% label for first subfigure
		\includegraphics[width=3.1in]{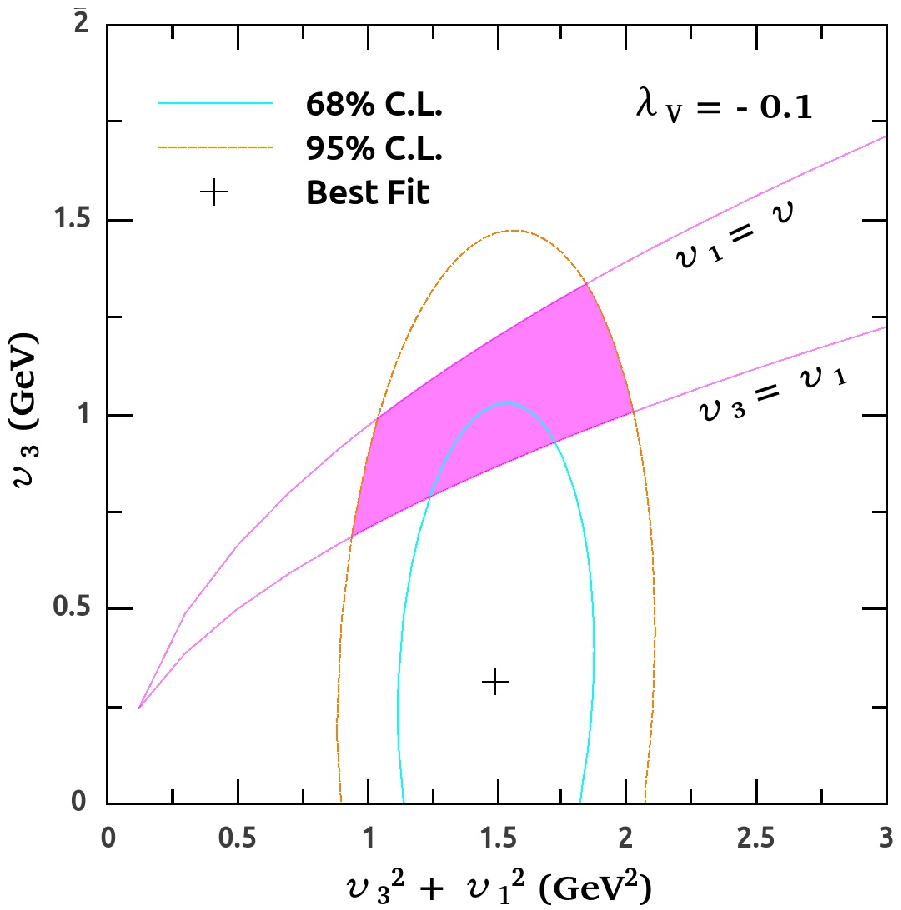}}
	\hspace{0.2in}
	\subfigure[]{
		\label{fig:subfig:b} %% label for second subfigure
		\includegraphics[width=3.0in]{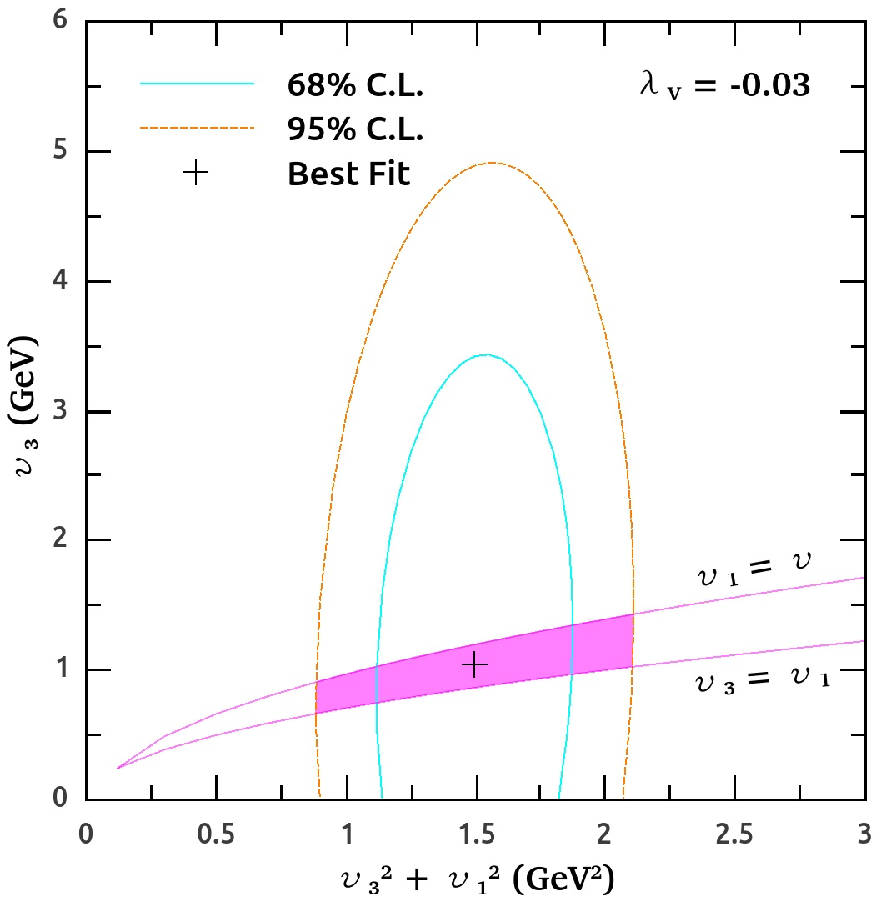}}
	\caption{ Limits on $\upsilon_3$ and $\upsilon_1$ from combination of Higgs fit and $B$-anomalies measurements as well as phenomenological condition, $\upsilon_3 > \upsilon_1> \upsilon$. The $U_1$ coupling is assumed with value of (a) $\lambda_V = - 0.1$ and (b) $\lambda_V = -0.03$. The LQ is survived in the pink regions.  \protect\label{fig:vvv}}
\end{figure}

Assuming $\lambda_{V} = -0.1 (-0.03)$, we show the constraints on $\upsilon_{3,1}$ in Fig.~\ref{fig:vvv}, which are obtained from combined limits of LHC Higgs data and $B$-anomalies as well as the condition $\upsilon_3 > \upsilon_1 > \upsilon$. The best value of $\upsilon_3$ obtained under the assumption of $\lambda_{V} = -0.1$ does not in the allowed region as shown in Fig.~\ref{fig:vvv} (a), while that does for $\lambda_{V} = -0.03$ (see Fig.~\ref{fig:vvv} (b)).

If the Higgs coupling and $R_{D^{(*)}}$ and $R_{K^{(*)}}$ precisely measured in the future, we can determine VEVs $\upsilon_3$ and $\upsilon_1$. For $\lambda_{V} = -0.03$, at the best fit value point, we obtain
\begin{eqnarray}
\upsilon_3 = 1.051 \ {\rm TeV} \ {\rm and} \ \upsilon_1 = 0.625 \ {\rm TeV}.
\end{eqnarray}
Equivalently, we obtain the mass of $U_1$
\begin{eqnarray}
m_U = \dfrac{1}{2} g_4 \sqrt{\upsilon_3^2 + \upsilon_1^2} = 2.140 {\rm TeV}
\end{eqnarray}

Then, one can determine or constrain other parameters in the model directly by using $\upsilon_3 = 1.051 \ {\rm GeV} \ {\rm and} \ \upsilon_1 = 0.625 \ {\rm GeV}$, or together with other constraints. For example, we can directly determine the masses of the other two new gauge particles $g'$ and $Z'$ in the model~\cite{DiLuzio:2018zxy}. Assuming $g_4 = 3.5$ and $g_3 = 1.07$ as well as $g_1 = 0.364$, we obtain
\begin{eqnarray}
m_{g'} &=& \sqrt{\dfrac{1}{2} (g^2_4 + g^2_3)\upsilon^2_3} = 2.72 \ \text{TeV} , \\
m_{Z'} &=& \sqrt{\dfrac{1}{4} (\dfrac{3}{2}g^2_4 + g^2_1)(\dfrac{1}{3}\upsilon^2_3 + v^2_1)} = 1.88 \ \text{TeV} .
\end{eqnarray}

Alternatively, once two of the three massive particles $U_1$, $g'$ and $Z'$ are found at the LHC or future colliders, one can use these masses together with $\xi_V$ to determine the $U_1$-Higgs coupling $\lambda_{V}$.

\section{Conclusions}
\label{sec:5}

$B$-anomalies may be a long-awaited new physical signal, and is discussed extensively as a hot topic. The good performance in explaining $B$-anomalies indicates that LQ may be discovered in the near future.

LQs with mass value of TeV-scale can considerably modify loop-induced Higgs processes, $ggF$ production and $h \to \gamma \gamma$ decay, which depending on the coupling size of LQ interactions with the Higgs boson. We study contributions of single scalar or vector LQ to loop-induced Higgs processes by analyzing current LHC Higgs data. Scalar LQs are studied model-independently while vector LQ, $U_1(\mathbf{3},\mathbf{1}, 2/3)$, is discussed in so called  the '4321'model.

Constraints on sizes of portal interactions, $\lambda_S = \lambda_S (\upsilon / m_{S})^2$, of all possible scalar LQs are obtained. Currently, the constraints for all scalar LQs are still too loose to acquire exact information for scalar LQ with mass of TeV scale, although accuracy of the result in this
analysis is more than doubled compared with previous one by analyzing Higgs data from LHC Run I~\cite{Dorsner:2016wpm}.

For vector LQ, $U_1(\mathbf{3},\mathbf{1}, 2/3)$, the size of interaction between $U_1$ and Higgs boson is parameterized as
\begin{eqnarray}
\xi_V = \dfrac{\lambda_V \upsilon \upsilon_3}{\upsilon^2_3 + \upsilon^2_1}, \nonumber
\end{eqnarray}
where $\lambda_V$ is the $U_1$-Higgs coupling constant. The best value with standard error and 95\% C.L. intervals for $\xi_V$ obtained from the Higgs fit read
\begin{eqnarray}
\xi_V = -0.005 \pm 0.008 , \ \xi_V \in \left[ -0.021, \ 0.011 \right]. \nonumber
\end{eqnarray}
The LQ coupling $\lambda_V$ is constrained to be small ($< 0.3$) for TeV-scale mass $U_1$, which is in accordance with the prediction of the '4321' model.

Compared to the present precision on $\xi_{S(V)}$, the situation is expected to improve by a factor of 2.4 at the HL-LHC.

We provide a method to determine VEVs, $\upsilon_3$ and $\upsilon_1$, of new scalar fields, $\Omega_3$ and $\Omega_1$ in the '4321'model, via the combination of the relation $\xi_V = \lambda_V \upsilon \upsilon_3/(\upsilon^2_3 + \upsilon^2_1)$ together with direct searches of $U_1$ at colliders as well as other constraints such as measurements of $R_{D^{(*)}}$ and $R_{K^{(*)}}$.

For conclusion, loop-induced Higgs processes $ggF$ production and $h \to \gamma \gamma$ decay are important processes that contributions of new heavy particles such as LQs may hidden in. We expect more precise measurements of Higgs properties in the future to guide us in the direction for LQ study.

\section*{Acknowledgements}
We thank Ilja Dor\v{s}ner for discussions. This work is supported in part by the National Natural Science Foundation of China under Grants No.11875157 and 11847303 (C. X. Yue and J. Zhang), and Liaoning Revitalization Talents Program No.60618009 (C. H. Li).

\appendix
\renewcommand{\appendixname}{Appendix~\Alph{section}}

\section{} \label{sec:A}

\subsection*{Contributions of $U_1$ to $R_{D^{(*)}}$ and $R_{K^{(*)}}$}

For the charged current $b \to c \ell \nu$, $U_1$ modifies $R_{D^{(*)}} =  \frac{\mathcal{B}(\bar{B} \to D^{(*)} \tau^- \bar{\nu})}{\mathcal{B}(\bar{B} \to D^{(*) \ell^- \bar{\nu})}}$ ($\ell = e, \mu $) by~\cite{DiLuzio:2018zxy}
\begin{eqnarray}
\Delta R_{D^{(*)}} &=& \frac{R^{\rm exp}_{D^{(*)}}}{R^{\rm SM}_{D^{(*)}}} -1 \nonumber \\
&\approx& 0.2 \left( \dfrac{2 \ {\rm TeV}}{m_U} \right)^2 \left( \frac{g_4}{3.5} \right)^2 {\rm sin} (2 \theta_{LQ}) \left( \frac{s_{\ell_3}}{0.8} \right)^2 \left( \frac{s_{q_3}}{0.8} \right) \left( \frac{s_{q_2}}{0.35} \right).
\end{eqnarray}
Setting $\theta_{LQ} = \pi/4$, $s_{\ell_3} = s_{q_3} = 0.8$, $s_{q_2} = 0.35$ and $g_4 = 3.5$, we obtain
\begin{eqnarray}
\label{eq:delRD}
\Delta R_{D^{(*)}} \approx 0.2 \left( \frac{2000 \ {\rm GeV}}{m_U} \right)^2 \approx 0.2 \dfrac{ \left( 1143 \ {\rm GeV} \right)^2}{ \upsilon^2_3 +\upsilon^2_1 }.
\end{eqnarray}

For neutral currents $b \to s \ell \ell$ case, $U_1$'s tree level contributions to Wilson coefficients $C^{\mu \mu}_9$ and $C^{\mu \mu}_{10}$ ($C_i = C^{SM}_i + \Delta C_i$) in the '4321' model are given by~\cite{DiLuzio:2018zxy}
\begin{eqnarray}
\left. \Delta C^{\mu \mu}_9 \right|_{\rm tree} = - \left. \Delta C^{\mu \mu}_{10} \right|_{\rm tree} = \frac{2 \pi}{\alpha_{em} V_{tb}V^*_{ts}} C_U \beta_{s \mu} \beta^*_{b \mu} ,
\end{eqnarray}
where $C_U = g^2_4 \upsilon^2 / (4 m^2_U)$, $\beta_{s \mu} = c_{\theta_{LQ}} s_{q_2} s_{\ell_2}$, $\beta_{b \mu} = -s_{\theta_{LQ}} s_{q_3} s_{\ell_2}$. For $\theta_{LQ} = \pi/4$, $s_{\ell_2} = 0.06$, $s_{q_3} = 0.8$, $s_{q_2} = 0.35$ and $g_4 = 3.5$, we obtain
\begin{eqnarray}
\label{eq:c9tree}
\left. \Delta C^{\mu \mu}_9 \right|_{\rm tree} = - \left. \Delta C^{\mu \mu}_{10} \right|_{\rm tree} = -0.46 \frac{\left( 1143 \ {\rm GeV} \right)^2}{\upsilon^2_3 + \upsilon^2_1}
\end{eqnarray}

One-loop log-enhanced processes at the scale of the bottom mass may  also contribute to the neutral currents sizeable. The contribution of the loops only to $C^{\ell \ell}_9$, which, in the $\beta_{b \tau}|V_{ts}|\ll \beta_{s \tau}$ limit, is given by~\cite{DiLuzio:2018zxy}
\begin{eqnarray}
\label{eq:c9loop}
\left. \Delta C^{\ell \ell}_9 \right|_{\rm loop} \left( m^2_b \right) \approx \frac{1}{3} \Delta R_{D^{(*)}} \left( {\rm log}x_b - \frac{1}{s^2_{\tau}} {\rm log}x_{E_2} \right),
\end{eqnarray}
where $x_{\alpha} = m^2_{\alpha}/m^2_{U}$, $E_2$ is a vector-like lepton introduced in the model. The contribution is universal for all leptons. Taking Eq.~(\ref{eq:delRD}) in to the above equation and setting $s_{\tau} = 0.8$, $m_{E_2} = 850$ GeV, we have
\begin{eqnarray}
\label{eq:delCll}
\left. \Delta C^{\ell \ell}_9 \right|_{\rm loop} \left( m^2_b \right) \approx \frac{0.2}{3} \dfrac{ \left(1143 \ {\rm GeV}\right)^2}{\upsilon^2_3 +\upsilon^2_1 }  \left( {\rm log} \frac{ \left(4.8 \times 10^{-3} \ {\rm TeV}\right)^2 }{ \upsilon^2_3 +\upsilon^2_1 } - \frac{1}{0.8^2} {\rm log}\frac{\left(0.97 \ {\rm TeV}\right)^2}{ \upsilon^2_3 +\upsilon^2_1 } \right).
\end{eqnarray}

Thus $U_1$ modifies the $b \to s \ell \ell$ processes via
\begin{eqnarray}
\delta C^{\mu \mu}_9 &=& \left. \Delta C^{\mu \mu}_{9, U_1} \right|_{\rm tree} + \left. \Delta C^{\mu \mu}_{9, U_1} \right|_{\rm loop}, \\
\delta C^{\mu \mu}_{10} &=& - \left. \Delta C^{\mu \mu}_{9, U_1} \right|_{\rm tree},\\
\delta C^{e e}_9 &=& \left. \Delta C^{\mu \mu}_9 \right|_{\rm loop}, \\
\delta C^{e e}_{10} &=& 0.
\end{eqnarray}

In this analysis, we consider the $U_1$ contributions to $b \to s \ell \ell$ processes in the case of

$\mathbf{scenario \ A}$. only via tree level contributions (Eq.~(\ref{eq:c9tree})), i.e.
\begin{eqnarray}
\delta C^{\mu \mu}_9 &=& - \delta C^{\mu \mu}_{10} = \left. \Delta C^{\mu \mu}_{9, U_1} \right|_{\rm tree}, \\ \nonumber
\delta C^{e e}_9 &=& \ ~\delta C^{e e}_{10} = 0.\nonumber
\end{eqnarray}

$\mathbf{scenario \ B}$. only via loop contributions (Eq.~(\ref{eq:c9loop})), i.e.
\begin{eqnarray}
\delta C^{\mu \mu}_9 &=& \delta C^{e e}_9 = \left. \Delta C^{\mu \mu}_{9, U_1} \right|_{\rm loop}, \\ \nonumber
\delta C^{\mu \mu}_{10} &=& \delta C^{e e}_{10} = 0.\nonumber
\end{eqnarray}

\subsection*{Fit to $R_{D^{(*)}}$ and $R_{K^{(*)}}$ measurements}

The newest average values of $R_{D}$ and $R_{D^{*}}$ including preliminary results at Belle II experiment~\cite{Abdesselam:2019dgh} are given by~\cite{Murgui:2019czp}
\begin{eqnarray}
R_{D} = 0.337 \pm 0.030 \quad  {\rm and} \quad  R_{D^{*}} = 0.299 \pm 0.013,
\end{eqnarray}
with a correlation of -0.36. The SM predictions of these two measurements read
\begin{eqnarray}
R^{SM}_{D} = 0.300^{+0.005}_{-0.004} \quad  {\rm and} \quad  R^{SM}_{D^{*}} = 0.251^{+0.004}_{-0.003}.
\end{eqnarray}
Then we obtain
\begin{eqnarray}
\label{eq:drde}
\Delta R_{D} = 0.123 \pm 0.101 \quad  {\rm and} \quad  \Delta R_{D^{*}} = 0.191\pm 0.054,
\end{eqnarray}
the correlation between the two quantities reads -0.34.

Ref.~\cite{Arbey:2019duh} has updated the $b \to s$ anomalies by including newest measurements of $R_{K}$ measured by the LHCb collaboration~\cite{Aaij:2019wad}, $R_{K^{*}}$ measured by the Belle
collaboration~\cite{Abdesselam:2019wac} as well as $B_{s,d} \to \mu^+ \mu^-$ measured by the ATLAS collaboration~\cite{Aaboud:2018mst}. The best fit values of $\left. \Delta C^{\mu \mu}_{9, U_1} \right|_{\rm tree}$  and $\left. \Delta C^{\mu \mu}_{9, U_1} \right|_{\rm loop}$ read respectively
\begin{eqnarray}
\label{eq:dc9}
\left. \Delta C^{\mu \mu}_{9, U_1} \right|_{\rm tree} = -0.41 \pm 0.10 \quad {\rm and} \quad \left. \Delta C_{9, U_1} \right|_{\rm loop} = -1.01 \pm 0.20.
\end{eqnarray}

To obtain $\upsilon^2_3 + \upsilon^2_1$, we perform our fit to measurements in Eqs.~(\ref{eq:drde}) and (\ref{eq:dc9}) by minimizing
\begin{eqnarray} \label{eq:xiB}
\chi^2_B = \left( \Delta R^{exp} -  \Delta R^{the} \right) C^{-1}_{\Delta R} \left( \Delta R^{exp} -  \Delta R^{the} \right) + \frac{\left(\Delta C^{exp}_{9} - \Delta C^{the}_{9} \right)^2}{ \left(\delta C^{\mu \mu}_{9}\right)^2},
\end{eqnarray}
where $\Delta R^{exp}$ denotes the measurement of $\Delta R_{D^{(*)}}$ and $\Delta R^{the}$ represents its theoretical prediction as shown in Eq.~{\ref{eq:delRD}}. Similarly, $\Delta C^{\mu \mu, exp}_{9}$ denotes $\Delta C^{\mu \mu}_{9}$ measured at experiments and $\Delta C^{\mu \mu, the}_{9}$ is its theoretical prediction as shown in Eq.~{\ref{eq:c9tree}} or Eq.~{\ref{eq:c9loop}}.

Then, we obtain the best fit value and preferred 95\% C.L. intervals of $\upsilon^2_3 + \upsilon^2_1$, for the case of scenario $\mathbf{A}$
\begin{eqnarray}
\upsilon^2_3 + \upsilon^2_1 &=& 1.496 \pm 0.250 \ {\rm TeV}^2, \\
\upsilon^2_3 + \upsilon^2_1  &\in& \left[ 1.127, 2.226 \right] \ {\rm TeV}^2 \ {\rm at \ 95\% C.L.},
\end{eqnarray}
for the case of scenario $\mathbf{B}$
\begin{eqnarray}
\upsilon^2_3 + \upsilon^2_1 &=& 1.220 \pm 0.187 \ {\rm TeV}^2, \\
\upsilon^2_3 + \upsilon^2_1  &\in& \left[ 0.939, 1.748 \right] \ {\rm TeV}^2 \ {\rm at \ 95\% C.L.}.
\end{eqnarray}

%%%%%%%%%%%%%%%%%%%%%%%%%%%%%%%%%%%%%%%%%%%%%%%%%%%%%%%%%%%%%%%%


\begin{thebibliography}{99}
	% b to s uu
	%\cite{Aaij:2014ora}
	\bibitem{Aaij:2014ora}
	R.~Aaij {\it et al.} [LHCb Collaboration],
	%``Test of lepton universality using $B^{+}\rightarrow K^{+}\ell^{+}\ell^{-}$ decays,''
	Phys.\ Rev.\ Lett.\  {\bf 113} (2014) 151601
	%doi:10.1103/PhysRevLett.113.151601
	[arXiv:1406.6482 [hep-ex]].
	
	%\cite{Aaij:2017vbb}
	\bibitem{Aaij:2017vbb}
	R.~Aaij {\it et al.} [LHCb Collaboration],
	%``Test of lepton universality with $B^{0} \rightarrow K^{*0}\ell^{+}\ell^{-}$ decays,''
	JHEP {\bf 1708} (2017) 055
	%doi:10.1007/JHEP08(2017)055
	[arXiv:1705.05802 [hep-ex]].
	
	%\cite{Bordone:2016gaq}
	\bibitem{Bordone:2016gaq}
	M.~Bordone, G.~Isidori and A.~Pattori,
	%``On the Standard Model predictions for $R_K$ and $R_{K^*}$,''
	Eur.\ Phys.\ J.\ C {\bf 76} (2016) no.8,  440
	%doi:10.1140/epjc/s10052-016-4274-7
	[arXiv:1605.07633 [hep-ph]].
	
	%\cite{Capdevila:2017ert}
	\bibitem{Capdevila:2017ert}
	B.~Capdevila, S.~Descotes-Genon, L.~Hofer and J.~Matias,
	%``Hadronic uncertainties in $B \to K^* \mu^+ \mu^-$: a state-of-the-art analysis,''
	JHEP {\bf 1704} (2017) 016
	%doi:10.1007/JHEP04(2017)016
	[arXiv:1701.08672 [hep-ph]].
	
	% b to c l v
	%\cite{Aaij:2015yra}
	\bibitem{Aaij:2015yra}
	R.~Aaij {\it et al.} [LHCb Collaboration],
	%``Measurement of the ratio of branching fractions $\mathcal{B}(\bar{B}^0 \to D^{*+}\tau^{-}\bar{\nu}_{\tau})/\mathcal{B}(\bar{B}^0 \to D^{*+}\mu^{-}\bar{\nu}_{\mu})$,''
	Phys.\ Rev.\ Lett.\  {\bf 115} (2015) no.11,  111803
	Erratum: [Phys.\ Rev.\ Lett.\  {\bf 115} (2015) no.15,  159901]
	%doi:10.1103/PhysRevLett.115.159901, 10.1103/PhysRevLett.115.111803
	[arXiv:1506.08614 [hep-ex]].
	
	%\cite{Huschle:2015rga}
	\bibitem{Huschle:2015rga}
	M.~Huschle {\it et al.} [Belle Collaboration],
	%``Measurement of the branching ratio of $\bar{B} \to D^{(\ast)} \tau^- \bar{\nu}_\tau$ relative to $\bar{B} \to D^{(\ast)} \ell^- \bar{\nu}_\ell$ decays with hadronic tagging at Belle,''
	Phys.\ Rev.\ D {\bf 92} (2015) no.7,  072014
	%doi:10.1103/PhysRevD.92.072014
	[arXiv:1507.03233 [hep-ex]].
	
	%\cite{Sato:2016svk}
	\bibitem{Sato:2016svk}
	Y.~Sato {\it et al.} [Belle Collaboration],
	%``Measurement of the branching ratio of $\bar{B}^0 \rightarrow D^{*+} \tau^- \bar{\nu}_{\tau}$ relative to $\bar{B}^0 \rightarrow D^{*+} \ell^- \bar{\nu}_{\ell}$ decays with a semileptonic tagging method,''
	Phys.\ Rev.\ D {\bf 94} (2016) no.7,  072007
	%doi:10.1103/PhysRevD.94.072007
	[arXiv:1607.07923 [hep-ex]].
	
	%\cite{Hirose:2016wfn}
	\bibitem{Hirose:2016wfn}
	S.~Hirose {\it et al.} [Belle Collaboration],
	%``Measurement of the $\tau$ lepton polarization and $R(D^*)$ in the decay $\bar{B} \to D^* \tau^- \bar{\nu}_\tau$,''
	Phys.\ Rev.\ Lett.\  {\bf 118} (2017) no.21,  211801
	%doi:10.1103/PhysRevLett.118.211801
	[arXiv:1612.00529 [hep-ex]].
	
	%\cite{Lees:2012xj}
	\bibitem{Lees:2012xj}
	J.~P.~Lees {\it et al.} [BaBar Collaboration],
	%``Evidence for an excess of $\bar{B} \to D^{(*)} \tau^-\bar{\nu}_\tau$ decays,''
	Phys.\ Rev.\ Lett.\  {\bf 109} (2012) 101802
	%doi:10.1103/PhysRevLett.109.101802
	[arXiv:1205.5442 [hep-ex]].
	
	%\cite{Lees:2013uzd}
	\bibitem{Lees:2013uzd}
	J.~P.~Lees {\it et al.} [BaBar Collaboration],
	%``Measurement of an Excess of $\bar{B} \to D^{(*)}\tau^- \bar{\nu}_\tau$ Decays and Implications for Charged Higgs Bosons,''
	Phys.\ Rev.\ D {\bf 88} (2013) no.7,  072012
	%doi:10.1103/PhysRevD.88.072012
	[arXiv:1303.0571 [hep-ex]].
	
	%\cite{HFAG2017}
	\bibitem{HFAG2017}
	Heavy Flavor Averaging Group, Average of $R_D$ and $R_D^*$ for FPCP 2017, http://www.slac.stanford.edu/xorg/hfag/semi/fpcp17/RDRDs.html (2017).
	
	%\cite{Fajfer:2012vx}
	\bibitem{Fajfer:2012vx}
	S.~Fajfer, J.~F.~Kamenik and I.~Nisandzic,
	%``On the $B \to D^* \tau \bar \nu_{\tau}$ Sensitivity to New Physics,''
	Phys.\ Rev.\ D {\bf 85} (2012) 094025
	%doi:10.1103/PhysRevD.85.094025
	[arXiv:1203.2654 [hep-ph]].
	
	%\cite{Aoki:2016frl}
	\bibitem{Aoki:2016frl}
	S.~Aoki {\it et al.},
	%``Review of lattice results concerning low-energy particle physics,''
	Eur.\ Phys.\ J.\ C {\bf 77} (2017) no.2,  112
	%doi:10.1140/epjc/s10052-016-4509-7
	[arXiv:1607.00299 [hep-lat]].
	%
	% leptoquark and B
	%
	%\cite{Gripaios:2014tna}
	\bibitem{Gripaios:2014tna}
	B.~Gripaios, M.~Nardecchia and S.~A.~Renner,
	%``Composite leptoquarks and anomalies in $B$-meson decays,''
	JHEP {\bf 1505} (2015) 006
	%doi:10.1007/JHEP05(2015)006
	[arXiv:1412.1791 [hep-ph]].
	
	%\cite{Georgi:2016xhm}
	\bibitem{Georgi:2016xhm}
	H.~Georgi and Y.~Nakai,
	%``Diphoton resonance from a new strong force,''
	Phys.\ Rev.\ D {\bf 94} (2016) no.7,  075005
	%doi:10.1103/PhysRevD.94.075005
	[arXiv:1606.05865 [hep-ph]].
	
	%\cite{Becirevic:2016yqi}
	\bibitem{Becirevic:2016yqi}
	D.~Bečirević, S.~Fajfer, N.~Košnik and O.~Sumensari,
	%``Leptoquark model to explain the $B$-physics anomalies, $R_K$ and $R_D$,''
	Phys.\ Rev.\ D {\bf 94} (2016) no.11,  115021
	%doi:10.1103/PhysRevD.94.115021
	[arXiv:1608.08501 [hep-ph]].
	
	%\cite{Becirevic:2017jtw}
	\bibitem{Becirevic:2017jtw}
	D.~Bečirević and O.~Sumensari,
	%``A leptoquark model to accommodate $R_K^\mathrm{exp} < R_K^\mathrm{SM}$ and $R_{K^\ast}^\mathrm{exp} < R_{K^\ast}^\mathrm{SM}$,''
	JHEP {\bf 1708} (2017) 104
	%doi:10.1007/JHEP08(2017)104
	[arXiv:1704.05835 [hep-ph]].
	
	%\cite{Diaz:2017lit}
	\bibitem{Diaz:2017lit}
	B.~Diaz, M.~Schmaltz and Y.~M.~Zhong,
	%``The leptoquark Hunter’s guide: Pair production,''
	JHEP {\bf 1710} (2017) 097
	%doi:10.1007/JHEP10(2017)097
	[arXiv:1706.05033 [hep-ph]].
	
	\bibitem{Buttazzo:2017ixm}
	D.~Buttazzo, A.~Greljo, G.~Isidori and D.~Marzocca,
	%``B-physics anomalies: a guide to combined explanations,''
	JHEP {\bf 1711} (2017) 044
	%doi:10.1007/JHEP11(2017)044
	[arXiv:1706.07808 [hep-ph]].
	
	%\cite{Guo:2017gxp}
	\bibitem{Guo:2017gxp}
	S.~Y.~Guo, Z.~L.~Han, B.~Li, Y.~Liao and X.~D.~Ma,
	%``Interpreting the $R_{K^{(*)}}$ anomaly in the colored Zee–Babu model,''
	Nucl.\ Phys.\ B {\bf 928} (2018) 435
	%doi:10.1016/j.nuclphysb.2018.01.024
	[arXiv:1707.00522 [hep-ph]].
	
	\bibitem{DiLuzio:2017vat}
	L.~Di Luzio, A.~Greljo and M.~Nardecchia,
	%``Gauge leptoquark as the origin of B-physics anomalies,''
	Phys.\ Rev.\ D {\bf 96} (2017) no.11,  115011
	%doi:10.1103/PhysRevD.96.115011
	[arXiv:1708.08450 [hep-ph]].
	
	%\cite{Calibbi:2017qbu}
	\bibitem{Calibbi:2017qbu}
	L.~Calibbi, A.~Crivellin and T.~Li,
	%``Model of vector leptoquarks in view of the $B$-physics anomalies,''
	Phys.\ Rev.\ D {\bf 98} (2018) no.11,  115002
	%doi:10.1103/PhysRevD.98.115002
	[arXiv:1709.00692 [hep-ph]].
	
	%\cite{Blanke:2018sro}
	\bibitem{Blanke:2018sro}
	M.~Blanke and A.~Crivellin,
	%``$B$ Meson Anomalies in a Pati-Salam Model within the Randall-Sundrum Background,''
	Phys.\ Rev.\ Lett.\  {\bf 121} (2018) no.1,  011801
	%doi:10.1103/PhysRevLett.121.011801
	[arXiv:1801.07256 [hep-ph]].
	
	%\cite{Fajfer:2018bfj}
	\bibitem{Fajfer:2018bfj}
	S.~Fajfer, N.~Košnik and L.~Vale Silva,
	%``Footprints of leptoquarks: from $ R_{K^{(*)}} $ to $ K \rightarrow \pi \nu \bar{\nu }$,''
	Eur.\ Phys.\ J.\ C {\bf 78} (2018) no.4,  275
	%doi:10.1140/epjc/s10052-018-5757-5
	[arXiv:1802.00786 [hep-ph]].
	
	%\cite{Matsuzaki:2018jui}
	\bibitem{Matsuzaki:2018jui}
	S.~Matsuzaki, K.~Nishiwaki and K.~Yamamoto,
	%``Simultaneous interpretation of $K$ and $B$ anomalies in terms of chiral-flavorful vectors,''
	JHEP {\bf 1811} (2018) 164
	%doi:10.1007/JHEP11(2018)164
	[arXiv:1806.02312 [hep-ph]].
	
	%\cite{Hati:2018fzc}
	\bibitem{Hati:2018fzc}
	C.~Hati, G.~Kumar, J.~Orloff and A.~M.~Teixeira,
	%``Reconciling $B$-meson decay anomalies with neutrino masses, dark matter and constraints from flavour violation,''
	JHEP {\bf 1811} (2018) 011
	%doi:10.1007/JHEP11(2018)011
	[arXiv:1806.10146 [hep-ph]].
	
	%\cite{Becirevic:2018afm}
	\bibitem{Becirevic:2018afm}
	D.~Bečirević, I.~Doršner, S.~Fajfer, N.~Košnik, D.~A.~Faroughy and O.~Sumensari,
	%``Scalar leptoquarks from grand unified theories to accommodate the $B$-physics anomalies,''
	Phys.\ Rev.\ D {\bf 98} (2018) no.5,  055003
	%doi:10.1103/PhysRevD.98.055003
	[arXiv:1806.05689 [hep-ph]].
	
	%\cite{Crivellin:2018yvo}
	\bibitem{Crivellin:2018yvo}
	A.~Crivellin, C.~Greub, D.~Müller and F.~Saturnino,
	%``Importance of Loop Effects in Explaining the Accumulated Evidence for New Physics in B Decays with a Vector Leptoquark,''
	Phys.\ Rev.\ Lett.\  {\bf 122} (2019) no.1,  011805
	%doi:10.1103/PhysRevLett.122.011805
	[arXiv:1807.02068 [hep-ph]].
	
	%\cite{deMedeirosVarzielas:2018bcy}
	\bibitem{deMedeirosVarzielas:2018bcy}
	I.~de Medeiros Varzielas and S.~F.~King,
	%``$ {R}_{K^{\left(*\right)}} $ with leptoquarks and the origin of Yukawa couplings,''
	JHEP {\bf 1811} (2018) 100
	%doi:10.1007/JHEP11(2018)100
	[arXiv:1807.06023 [hep-ph]].
	
	%\cite{Azatov:2018kzb}
	\bibitem{Azatov:2018kzb}
	A.~Azatov, D.~Barducci, D.~Ghosh, D.~Marzocca and L.~Ubaldi,
	%``Combined explanations of B-physics anomalies: the sterile neutrino solution,''
	JHEP {\bf 1810} (2018) 092
	%doi:10.1007/JHEP10(2018)092
	[arXiv:1807.10745 [hep-ph]].
	
	\bibitem{DiLuzio:2018zxy}
	L.~Di Luzio, J.~Fuentes-Martin, A.~Greljo, M.~Nardecchia and S.~Renner,
	%``Maximal Flavour Violation: a Cabibbo mechanism for leptoquarks,''
	JHEP {\bf 1811} (2018) 081
	%doi:10.1007/JHEP11(2018)081
	[arXiv:1808.00942 [hep-ph]].
	
	%\cite{Faber:2018qon}
	\bibitem{Faber:2018qon}
	T.~Faber, M.~Hudec, M.~Malinský, P.~Meinzinger, W.~Porod and F.~Staub,
	%``A unified leptoquark model confronted with lepton non-universality in $B$-meson decays,''
	Phys.\ Lett.\ B {\bf 787} (2018) 159
	%doi:10.1016/j.physletb.2018.10.051
	[arXiv:1808.05511 [hep-ph]].
	
	%\cite{Heeck:2018ntp}
	\bibitem{Heeck:2018ntp}
	J.~Heeck and D.~Teresi,
	%``Pati-Salam explanations of the B-meson anomalies,''
	JHEP {\bf 1812} (2018) 103
	%doi:10.1007/JHEP12(2018)103
	[arXiv:1808.07492 [hep-ph]].
	
	%\cite{Angelescu:2018tyl}
	\bibitem{Angelescu:2018tyl}
	A.~Angelescu, D.~Bečirević, D.~A.~Faroughy and O.~Sumensari,
	%``Closing the window on single leptoquark solutions to the $B$-physics anomalies,''
	JHEP {\bf 1810} (2018) 183
	%doi:10.1007/JHEP10(2018)183
	[arXiv:1808.08179 [hep-ph]].
	
	%\cite{Balaji:2018zna}
	\bibitem{Balaji:2018zna}
	S.~Balaji, R.~Foot and M.~A.~Schmidt,
	%``Chiral SU(4) explanation of the $b\to s$ anomalies,''
	Phys.\ Rev.\ D {\bf 99} (2019) no.1,  015029
	%doi:10.1103/PhysRevD.99.015029
	[arXiv:1809.07562 [hep-ph]].
	
	%\cite{Watanabe:2018jhh}
	\bibitem{Watanabe:2018jhh}
	R.~Watanabe,
	%``$R_D$ and $R_{D^*}$: Theoretical Development,''
	arXiv:1810.00379 [hep-ph].
	
	%\cite{Schmaltz:2018nls}
	\bibitem{Schmaltz:2018nls}
	M.~Schmaltz and Y.~M.~Zhong,
	%``The leptoquark Hunter’s guide: large coupling,''
	JHEP {\bf 1901} (2019) 132
	%doi:10.1007/JHEP01(2019)132
	[arXiv:1810.10017 [hep-ph]].
	
	%\cite{Bansal:2018nwp}
	\bibitem{Bansal:2018nwp}
	S.~Bansal, R.~M.~Capdevilla and C.~Kolda,
	%``Constraining the minimal flavor violating leptoquark explanation of the $R_{D^{(*)}}$  anomaly,''
	Phys.\ Rev.\ D {\bf 99} (2019) no.3,  035047
	%doi:10.1103/PhysRevD.99.035047
	[arXiv:1810.11588 [hep-ph]].
	
	%\cite{Iguro:2018vqb}
	\bibitem{Iguro:2018vqb}
	S.~Iguro, T.~Kitahara, Y.~Omura, R.~Watanabe and K.~Yamamoto,
	%``D$^{*}$ polarization vs. $ {R}_{D^{\left(\ast \right)}} $ anomalies in the leptoquark models,''
	JHEP {\bf 1902} (2019) 194
	%doi:10.1007/JHEP02(2019)194
	[arXiv:1811.08899 [hep-ph]].
	
	%\cite{Fajfer:2018hbq}
	\bibitem{Fajfer:2018hbq}
	S.~Fajfer,
	%``Scalar leptoquarks: From GUT to B anomalies,''
	EPJ Web Conf.\  {\bf 192} (2018) 00025.
	%doi:10.1051/epjconf/201819200025
	
	%\cite{Fornal:2018dqn}
	\bibitem{Fornal:2018dqn}
	B.~Fornal, S.~A.~Gadam and B.~Grinstein,
	%``Left-Right SU(4) Vector Leptoquark Model for Flavor Anomalies,''
	Phys.\ Rev.\ D {\bf 99} (2019) no.5,  055025
	%doi:10.1103/PhysRevD.99.055025
	[arXiv:1812.01603 [hep-ph]].
	
	%\cite{DaRold:2018moy}
	\bibitem{DaRold:2018moy}
	L.~Da Rold and F.~Lamagna,
	%``Composite Higgs and leptoquarks from a simple group,''
	JHEP {\bf 1903} (2019) 135
	%doi:10.1007/JHEP03(2019)135
	[arXiv:1812.08678 [hep-ph]].
	
	%\cite{deMedeirosVarzielas:2019lgb}
	\bibitem{deMedeirosVarzielas:2019lgb}
	I.~de Medeiros Varzielas and J.~Talbert,
	%``Simplified Models of Flavourful Leptoquarks,''
	arXiv:1901.10484 [hep-ph].
	
	%\cite{Zhang:2019hth}
	\bibitem{Zhang:2019hth}
	J.~Zhang, Y.~Zhang, Q.~Zeng and R.~Sun,
	%``New physics effects of the vector leptoquark on ${\bar{B}}^{*}\rightarrow P\tau {\bar{\nu }}_{\tau }$ decays,''
	Eur.\ Phys.\ J.\ C {\bf 79} (2019) no.2,  164.
	
	%\cite{Aydemir:2019ynb}
	\bibitem{Aydemir:2019ynb}
	U.~Aydemir, T.~Mandal and S.~Mitra,
	%``A single TeV-scale scalar leptoquark in $\mathbf{SO(10)}$ grand unification and $\mathbf{B}$-decay anomalies,''
	arXiv:1902.08108 [hep-ph].
	
	%\cite{Cata:2019wbu}
	\bibitem{Cata:2019wbu}
	O.~Catà and T.~Mannel,
	%``Linking lepton number violation with $B$ anomalies,''
	arXiv:1903.01799 [hep-ph].
	
	%\cite{Bhattacharya:2019olg}
	\bibitem{Bhattacharya:2019olg}
	B.~Bhattacharya, A.~Datta, S.~Kamali and D.~London,
	%``CP Violation in ${\bar B}^0\to D^{*+}\mu^-{\bar\nu}_\mu$,''
	arXiv:1903.02567 [hep-ph].
	
	%\cite{Adam:2019oes}
	\bibitem{Adam:2019oes}
	A.~S.~Adam, A.~Ferdiyan and M.~Satriawan,
	%``A New Left-Right Symmetry Model,''
	arXiv:1903.03370 [hep-ph].
	
	%\cite{Aebischer:2019mlg}
	\bibitem{Aebischer:2019mlg}
	J.~Aebischer, W.~Altmannshofer, D.~Guadagnoli, M.~Reboud, P.~Stangl and D.~M.~Straub,
	%``$B$-decay discrepancies after Moriond 2019,''
	arXiv:1903.10434 [hep-ph].
	
	%\cite{Cornella:2019hct}
	\bibitem{Cornella:2019hct}
	C.~Cornella, J.~Fuentes-Martin and G.~Isidori,
	%``Revisiting the vector leptoquark explanation of the B-physics anomalies,''
	arXiv:1903.11517 [hep-ph].
	
	%\cite{Barbieri:2019zdz}
	\bibitem{Barbieri:2019zdz}
	R.~Barbieri and R.~Ziegler,
	%``Quark masses, CKM angles and Lepton Flavour Universality violation,''
	arXiv:1904.04121 [hep-ph].
	
	
	
	%
	% leptoquark
	%
	\bibitem{Buchmuller:1986zs}
	W.~Buchmuller, R.~Ruckl and D.~Wyler,
	%``Leptoquarks in Lepton - Quark Collisions,''
	Phys.\ Lett.\ B {\bf 191} (1987) 442
	Erratum: [Phys.\ Lett.\ B {\bf 448} (1999) 320].
	%doi:10.1016/S0370-2693(99)00014-3, 10.1016/0370-2693(87)90637-X
	
	\bibitem{Dorsner:2016wpm}
	I.~Doršner, S.~Fajfer, A.~Greljo, J.~F.~Kamenik and N.~Košnik,
	%``Physics of leptoquarks in precision experiments and at particle colliders,''
	Phys.\ Rept.\  {\bf 641} (2016) 1
	%doi:10.1016/j.physrep.2016.06.001
	[arXiv:1603.04993 [hep-ph]].
	
	%\cite{Tanabashi:2018oca}
	\bibitem{Tanabashi:2018oca}
	M.~Tanabashi {\it et al.} [Particle Data Group],
	%``Review of Particle Physics,''
	Phys.\ Rev.\ D {\bf 98} (2018) no.3,  030001.
		
	\bibitem{Pati-Salam}
	J.~C.~Pati and A.~Salam,
	%``Lepton Number as the Fourth Color,''
	Phys.\ Rev.\ D {\bf 10} (1974) 275.
	Erratum: [Phys.\ Rev.\ D {\bf 11} (1975) 703].
	
	\bibitem{Georgi:1974sy}
	H.~Georgi and S.~L.~Glashow,
	%``Unity of All Elementary Particle Forces,''
	Phys.\ Rev.\ Lett.\  {\bf 32} (1974) 438.
	%doi:10.1103/PhysRevLett.32.438.
	
	\bibitem{Georgi:1974my}
	H.~Georgi,
	%``The State of the Art—Gauge Theories,''
	AIP Conf.\ Proc.\  {\bf 23} (1975) 575.
	%doi:10.1063/1.2947450
	
	\bibitem{technicolor}
	B.~Schrempp and F.~Schrempp,
	%``Light Leptoquarks,''
	Phys.\ Lett.\  {\bf 153B}, 101 (1985).	
		
	\bibitem{composite}
	J.~Wudka,
	%``Composite Leptoquarks,''
	Phys.\ Lett.\  {\bf 167B} (1986) 337.
	
	\bibitem{Dorsner:2015mja}
	I.~Doršner, S.~Fajfer, A.~Greljo, J.~F.~Kamenik, N.~Košnik and I.~Nišandžic,
	%``New Physics Models Facing Lepton Flavor Violating Higgs Decays at the Percent Level,''
	JHEP {\bf 1506} (2015) 108
	%doi:10.1007/JHEP06(2015)108
	[arXiv:1502.07784 [hep-ph]].
	
	\bibitem{Chang:2012ta}
	W.~F.~Chang, J.~N.~Ng and J.~M.~S.~Wu,
	%``Constraints on New Scalars from the LHC 125 GeV Higgs Signal,''
	Phys.\ Rev.\ D {\bf 86} (2012) 033003
	%doi:10.1103/PhysRevD.86.033003
	[arXiv:1206.5047 [hep-ph]].
	
	\bibitem{atlas-higgs}
	G.~Aad {\it et al.} [ATLAS Collaboration],
	%``The ATLAS Experiment at the CERN Large Hadron Collider,''
	JINST {\bf 3} (2008) S08003.

	\bibitem{cms-higgs}
	S.~Chatrchyan {\it et al.} [CMS Collaboration],
	%``The CMS Experiment at the CERN LHC,''
	JINST {\bf 3} (2008) S08004.
	
	\bibitem{Khachatryan:2016vau}
	G.~Aad {\it et al.} [ATLAS and CMS Collaborations],
	%``Measurements of the Higgs boson production and decay rates and constraints on its couplings from a combined ATLAS and CMS analysis of the LHC pp collision data at $ \sqrt{s}=7 $ and 8 TeV,''
	JHEP {\bf 1608} (2016) 045
	%doi:10.1007/JHEP08(2016)045
	[arXiv:1606.02266 [hep-ex]].
	
	%\cite{ATLAS:2019slw}
	\bibitem{ATLAS:2019slw}
	The ATLAS collaboration [ATLAS Collaboration],
	%``Combined measurements of Higgs boson production and decay using up to $80$ fb$^{-1}$ of proton--proton collision data at $\sqrt{s}=$ 13 TeV collected with the ATLAS experiment,''
	ATLAS-CONF-2019-005.
	
	%\cite{Sirunyan:2018koj}
	\bibitem{Sirunyan:2018koj}
	A.~M.~Sirunyan {\it et al.} [CMS Collaboration],
	%``Combined measurements of Higgs boson couplings in proton-proton collisions at $\sqrt{s}=$ 13 TeV,''
	Submitted to: Eur.Phys.J.
	[arXiv:1809.10733 [hep-ex]].
	
	\bibitem{TheATLASandCMSCollaborations:2015bln}
	The ATLAS and CMS Collaborations,
	%``Measurements of the Higgs boson production and decay rates and constraints on its couplings from a combined ATLAS and CMS analysis of the LHC pp collision data at $\sqrt{s}$ = 7 and 8 TeV,''
	ATLAS-CONF-2015-044.
	
	\bibitem{CMS:2015kwa}
	CMS Collaboration [CMS Collaboration],
	%``Measurements of the Higgs boson production and decay rates and constraints on its couplings from a combined ATLAS and CMS analysis of the LHC pp collision data at sqrt s = 7 and 8 TeV,''
	CMS-PAS-HIG-15-002.
	
	%\cite{Heinemeyer:2013tqa}
	\bibitem{Heinemeyer:2013tqa}
	S.~Heinemeyer {\it et al.} [LHC Higgs Cross Section Working Group],
	%``Handbook of LHC Higgs Cross Sections: 3. Higgs Properties,''
	%doi:10.5170/CERN-2013-004
	arXiv:1307.1347 [hep-ph].
		
	\bibitem{Enkhbat:2013oba}
	T.~Enkhbat,
	%``Scalar leptoquarks and Higgs pair production at the LHC,''
	JHEP {\bf 1401} (2014) 158
	%doi:10.1007/JHEP01(2014)158
	[arXiv:1311.4445 [hep-ph]].
		
	\bibitem{Djouadi:2005gj}
	A.~Djouadi,
	%``The Anatomy of electro-weak symmetry breaking. II. The Higgs bosons in the minimal supersymmetric model,''
	Phys.\ Rept.\  {\bf 459} (2008) 1
	%doi:10.1016/j.physrep.2007.10.005
	[hep-ph/0503173].
	
	\bibitem{Carena:2012xa}
	M.~Carena, I.~Low and C.~E.~M.~Wagner,
	%``Implications of a Modified Higgs to Diphoton Decay Width,''
	JHEP {\bf 1208} (2012) 060
	%doi:10.1007/JHEP08(2012)060
	[arXiv:1206.1082 [hep-ph]].	
	
	\bibitem{Dorsner:2012pp}
	I.~Dorsner, S.~Fajfer, A.~Greljo and J.~F.~Kamenik,
	%``Higgs Uncovering Light Scalar Remnants of High Scale Matter Unification,''
	JHEP {\bf 1211} (2012) 130
	%doi:10.1007/JHEP11(2012)130
	[arXiv:1208.1266 [hep-ph]].
	
	\bibitem{Agrawal:1999bk}
	P.~Agrawal and U.~Mahanta,
	%``Leptoquark contribution to the Higgs boson production at the CERN LHC collider,''
	Phys.\ Rev.\ D {\bf 61} (2000) 077701
	%doi:10.1103/PhysRevD.61.077701
	[hep-ph/9911497].
	
	\bibitem{Gori:2013mia}
	S.~Gori and I.~Low,
	%``Precision Higgs Measurements: Constraints from New Oblique Corrections,''
	JHEP {\bf 1309} (2013) 151
	%doi:10.1007/JHEP09(2013)151
	[arXiv:1307.0496 [hep-ph]].
	
	%\cite{Aaboud:2019jcc}
	\bibitem{Aaboud:2019jcc}
	M.~Aaboud {\it et al.} [ATLAS Collaboration],
	%``Searches for scalar leptoquarks and differential cross-section measurements in dilepton-dijet events in proton-proton collisions at a centre-of-mass energy of $\sqrt{s}$ = 13 TeV with the ATLAS experiment,''
	arXiv:1902.00377 [hep-ex].
	
	%\cite{Cepeda:2019klc}
	\bibitem{Cepeda:2019klc}
	M.~Cepeda {\it et al.} [Physics of the HL-LHC Working Group],
	%``Higgs Physics at the HL-LHC and HE-LHC,''
	arXiv:1902.00134 [hep-ph].
	
	%\cite{Sirunyan:2018kzh}
	\bibitem{Sirunyan:2018kzh}
	A.~M.~Sirunyan {\it et al.} [CMS Collaboration],
	%``Constraints on models of scalar and vector leptoquarks decaying to a quark and a neutrino at $\sqrt{s}=$ 13 TeV,''
	Phys.\ Rev.\ D {\bf 98} (2018) no.3,  032005
	%doi:10.1103/PhysRevD.98.032005
	[arXiv:1805.10228 [hep-ex]].
	
	%\cite{Murgui:2019czp}
	\bibitem{Murgui:2019czp}
	C.~Murgui, A.~Peñuelas, M.~Jung and A.~Pich,
	%``Global fit to $b \to c \tau \nu$ transitions,''
	arXiv:1904.09311 [hep-ph].
	
	%\cite{Abdesselam:2019dgh}
	\bibitem{Abdesselam:2019dgh}
	A.~Abdesselam {\it et al.} [Belle Collaboration],
	%``Measurement of $\mathcal{R}(D)$ and $\mathcal{R}(D^{\ast})$ with a semileptonic tagging method,''
	arXiv:1904.08794 [hep-ex].
	
	%\cite{Arbey:2019duh}
	\bibitem{Arbey:2019duh}
	A.~Arbey, T.~Hurth, F.~Mahmoudi, D.~Martinez Santos and S.~Neshatpour,
	%``Update on the b->s anomalies,''
	arXiv:1904.08399 [hep-ph].
	
	%\cite{Aaij:2019wad}
	\bibitem{Aaij:2019wad}
	R.~Aaij {\it et al.} [LHCb Collaboration],
	%``Search for lepton-universality violation in $B^+\to K^+\ell^+\ell^-$ decays,''
	arXiv:1903.09252 [hep-ex].
	
	%\cite{Abdesselam:2019wac}
	\bibitem{Abdesselam:2019wac}
	A.~Abdesselam {\it et al.} [Belle Collaboration],
	%``Test of lepton flavor universality in ${B\to K^\ast\ell^+\ell^-}$ decays at Belle,''
	arXiv:1904.02440 [hep-ex].
	
	%\cite{Aaboud:2018mst}
	\bibitem{Aaboud:2018mst}
	M.~Aaboud {\it et al.} [ATLAS Collaboration],
	%``Study of the rare decays of $B^0_s$ and $B^0$ mesons into muon pairs using data collected during 2015 and 2016 with the ATLAS detector,''
	JHEP {\bf 1904} (2019) 098
	doi:10.1007/JHEP04(2019)098
	[arXiv:1812.03017 [hep-ex]].

	
\end{thebibliography}
\end{document}